\documentclass[11pt]{article}
\usepackage{times}
\usepackage{a4}
\usepackage{amsfonts}
\usepackage{amsthm}
\usepackage{amsmath}
\usepackage{eufrak}
\usepackage{pstricks}
\usepackage{pst-node}
\usepackage[TS1,T1]{fontenc}
\usepackage[latin1]{inputenc}
\usepackage{mathrsfs}
\usepackage{booktabs}
\usepackage{fullpage}
\usepackage{cite}

\usepackage[all]{xy} 

\newbox\ncintdbox \newbox\ncinttbox 
\setbox0=\hbox{$-$}
\setbox2=\hbox{$\displaystyle\int$}
\setbox\ncintdbox=\hbox{\rlap{\hbox
    to \wd2{\hskip-.125em\box2\relax\hfil}}\box0\kern.1em}
\setbox0=\hbox{$\vcenter{\hrule width 4pt}$}
\setbox2=\hbox{$\textstyle\int$}
\setbox\ncinttbox=\hbox{\rlap{\hbox
    to \wd2{\hskip-.125em\box2\relax\hfil}}\box0\kern.1em}

\newcommand{\stroke}{\mathbin|}     

\newtheorem{lemma}{Lemma}[section]
\newtheorem{lema}[lemma]{Lemma}
\newtheorem{thm}[lemma]{Theorem}

\theoremstyle{definition}
\newtheorem{example}[lemma]{Example}
\newtheorem{defn}[lemma]{Definition}

\newcommand{\C}{\mathbb{C}}       
\renewcommand{\det}{\operatorname{det}} 
\DeclareMathOperator{\id}{id}     
\newcommand{\R}{\mathbb{R}}       
\DeclareMathOperator{\sign}{sign} 
\DeclareMathOperator{\tr}{tr}     

\newcommand{\bea}{\begin{eqnarray}}  
\newcommand{\eea}{\end{eqnarray}}  
\newcommand{\beas}{\begin{eqnarray*}} 
\newcommand{\eeas}{\end{eqnarray*}}  
\newcommand{\be}{\begin{equation}} 
\newcommand{\ee}{\end{equation}} 

\def\<#1,#2>{\langle#1\stroke#2\rangle} 

\def\C{{\mathbb C}}
\def\R{{\mathbb R}}

\def\CA{{\cal A}}

\def\CH{{\mathscr{H}}}

\def\CD{{\mathscr{D}}}

\def\id{\hbox{id}}

\begin{document}

\begin{flushright} 
MZ-TH/01-35
\end{flushright}  

\vspace{1cm} 

\begin{center} 
{\bf \Large Moduli spaces of discrete gravity I: A few points...} 

\vspace{1cm} 

{\large Alexander Holfter\footnote{E-mail: holfter@thep.physik.uni-mainz.de} and  
Mario Paschke\footnote{E-mail: paschke@thep.physik.uni-mainz.de \\ \\
Address: Institut f\"ur Physik, Universit\"at
  Mainz,  55099 Mainz, Germany}}
\end{center} 

\vspace{1cm} 

\begin{abstract}
  Spectral triples describe and generalize Riemannian spin
  geometries by converting the
  geometrical information into algebraic data, which consist of an
  algebra $A$, a Hilbert space $H$ carrying a representation of $A$ and
  the Dirac operator $D$ (a selfadjoint operator acting on $H$). 
  The gravitational action is described by the trace of a suitable
  function of $D$.

  In this paper we examine the (path-integral-) quantization of such a system given by a finite
  dimensional commutative algebra. 
  It is then (in concrete examples)  possible to construct the moduli space
  of the theory, i.e. to divide the space of all Dirac operators by the action of 
  the diffeomorphism group, and to construct an invariant measure 
  on this space.  
  
  We discuss expectation values of various 
  observables and demonstrate some interesting effects such as the
  effect of coupling the system to Fermions (which renders finite quantities in
  cases, where the Bosons alone would give infinite quantities) or the striking
  effect of spontaneous breaking of spectral invariance.

\end{abstract}

\section{Introduction}
Locally all manifolds look just like $\R^d$ and,  within the present experimental accuracy,
spacetime looks the same. ``But now it seems that the empirical notions, on which the measurement of distances is based,
namely the notion of a rigid body and a light ray, lose their validity at unmeasurable small scales.
Hence, it is well conceivable, that spacetime does not fulfill the suppositions of this (Riemannian) geometry,
and this one should in fact assume, if it would explain  observations more naturally.'' As has been pointed
out by B.Riemann 1854 in his famous habilitation thesis.

\noindent
Nowadays, it seems to be a common belief among theoretical physicists that spacetime possesses  (at 
small distances) a structure different from that of a classical manifold. There are, for instance, many    
heuristic arguments (e.g. \cite{doplicheretal}, or the ones cited in \cite{amelino}) for some uncertainty 
relation of the type  $[x^\mu , x^\nu ] \neq 0 $  caused by quantum effects in the process of measuring 
a spacetime region with a certain accuracy. Another appealing hint to a noncommutativity
of spacetime -- visible at high energies -- is Connes' and Chamseddines' \cite{spectralaction} description of the standard model 
of particle physics as a part of the gravitational field over an appropriately chosen
noncommutative space. Other approaches \cite{CES,CL,D-VKM2} formulate the standard model as gauge theory
over (essentially) the same noncommutative spacetime, 
which is given by the tensor product of the algebra of functions over spacetime with some finite dimensional $C^*$-algebra.

\noindent
In this paper will shall restrict ourselves to Connes' approach,
which describes the geometrical structure of spacetime, and especially its spin structure,
as a spectral triple $(\CA,\CH, D)$ \cite{polaris,connes,landi,gravitymatter}. 
The gravitational action can then, for instance, be reformulated by using the ``spectral action principle'' 
introduced in \cite{spectralaction} as the trace of a suitable function of the Dirac operator $D$. 
Hence the gravitational action is directly expressed as a sum over 
diffeomorphism invariant quantitities, namely the eigenvalues of the Dirac operator. 
Such an action is, of course, invariant not only under diffeomorphisms of spacetime, but 
under {\em all} unitaries on the Hilbert space $\CH$. That is, however,  too much symmetry
to simply reproduce the Einstein-Hilbert action of general relativity, since this action
is invariant under nothing but the group of diffeomorphisms. The reason for this increase
of symmetry is the fact that the Connes-Chamseddine action does not take into account boundary conditions 
for the metric (i.e. initial data). It only reproduces the integral over the {\em entire} spacetime
of the cosmological constant, the scalar curvature (and possibly higher order terms) \cite{diss}.
In fact, the eigenvalues of $D$ do not form a complete set of observables
for the gravitational field (``one cannot hear the shape of drum''). We refer to \cite{LaRo} for an instructive argument. \\
Finally, it should also be noted that this description only works for spaces with Euclidean
signature of the metric. The generalization to Lorentzian or
arbitrary signature is in progress
\cite{strohmaier,galatea,quadrupel}. The approach initiated in \cite{quadrupel} is actually designed to facilitate the incorporation 
of initial conditions for the metric in the theory.\\ 
Ignoring these open problems in the following, we shall focus our
attention on the ``established'' form of spectral triples with strict
Euclidean metric. 
The spectral action principle then defines an action for the metric corresponding to spectral triples,
that possesses all the required properties. 
It should be stressed that this theory -- despite its quantum like appearance -- is completely classical, and
this raises the problem of its quantization.\\
No need to say that this would be a rather hard task for generic spectral triples. However, for finite spectral
triples, for which the Hilbert space $\CH$ is finite dimensional, one can hope to undertake this job
and thereby gain some new insights that might be helpful also for the generic case. Apart from that, the study of finite
dimensional spectral quantum gravity might also be interesting by itself, as it gives rise to very unusual matrix models, with unusual symmetries
and a completely new ``physical'' interpretation.\\
In \cite{Rov} the canonical quantization of a particular example has been performed by making use of the observation that there
exists a canonical symplectic form on the space of Dirac operators.
That idea, however, only works as long as one ignores the (spectral or at least diffeomorphism) invariance
of the system, since only then the configuration space is simply given as the space of all Dirac operators.
In this series of papers we shall therefore address the problem of correct treatment of
the diffeomorphism invariance of spectral actions.

At first thought, it seems to be quite easy to formulate a path integral
for a spectral invariant system:\\
Denoting the independent eigenvalues of the Dirac operator by $\lambda_1 ,\ldots , \lambda_n$, any
spectral invariant measure can be written in the form
\[ {\rm d} \lambda_1 \cdots {\rm d} \lambda_n\,\,\,  \rho(\lambda_1,\ldots , \lambda_n ) \]
with a density $\rho$ that is symmetric under permutations of its arguments. 
(There do always exist unitaries which interchange the eigenspaces of D.)
Unfortunately it is usually inconvenient to express meaningful observables in terms of the eigenvalues of $D$, and
in fact, even for finite dimensional spectral triples some observables cannot be written that way.
For instance, for the two-point space that we consider in this first paper,
the distance of the two points is invariant under the group of diffeomorphisms -- which of course is not a very impressing property, as there
is only one nontrivial diffeomorphism: the interchange of the two points -- and hence an observable.
We shall present an example where it is functionally independent of the eigenvalues of the Dirac operator.\\
So (for such examples) one would only seek for a diffeomorphism invariant measure and  try to write it using the entries of $D$.
In \cite{MPendlicheTripel,TKendlicheTripel} a classification of finite dimensional spectral
triples has been given, which includes an explicit characterization of the space of all
Dirac operators or, more precisely, of those entries which do not necessarily vanish.
There did remain, however, a freedom of choice of basis in $\CH$, which allows to restrict the range of some entries or
even transform them to zero.\\
Two spectral triples are called unitarily equivalent if they can be transformed into each other by such a choice of basis
in $\CH$, i.e. if all the data of the triples can be transformed into each other by the same unitary transformation $U$.
Apart from the Dirac operator, it is not hard to classify 
all data of real, even, finite dimensional spectral triples -- i.e. the representations $\pi, \pi^o$ of the algebra $\CA$,
the grading $\Gamma$ and the reality structure $J$ -- up to unitary equivalence. That has been done in  
\cite{MPendlicheTripel,TKendlicheTripel}, we shall shortly review the results in section \ref{review}.\\
But then there still exist unitaries on $\CH$, which do commute with $\pi, \pi^o,\Gamma$ and $J$ but not with the Dirac operator.
Presently it is unfortunately not possible to systematically investigate the consequences of this additional
freedom for the classification of 
finite dimensional spectral triples {\em up to unitary equivalence}. 
For particular examples one can, however, divide the space of Dirac operators by the action of all (such) unitaries.  
We shall present some of these examples in 
section \ref{moduli}.  \\
\\
In the path integral quantization of spectral triples one should, of course, only sum over unitary equivalence classes, and hence
the computation of these moduli spaces is essential for our project.
As mentioned above, in this paper we shall concentrate on concrete examples based on the two-point space $\CA=\C^2$.
We hope that they can serve to illustrate not only the complexity of the problem, but also that its study is not in vain.
We plan to give a more systematic construction of path integrals for finite spectral triples in the third paper of this series.\\
Note that the diffeomorphisms of the underlying spacetime, i.e. the automorphisms of $\CA$, do not belong to
the unitary equivalences that we considered above: If they are actually represented unitarily on $\CH$, they do not commute with the
representations of $\CA$. So, in the generic case, one should in addition divide the space of Dirac operators by the action
of the diffeomorphism group. In the $\C^2$-examples which we present in this paper, this is, however trivial.
In the next paper we shall investigate a truely noncommutative example based on the algebra $M_2(\C)\oplus \C$ where the diffeomorphisms
play an important role.\\ \\
Despite their simplicity, our examples show several interesting features.\\
At first view, the most striking effect that we find seems to be the spontaneous
breaking of spectral invariance, which is actually rather easily understood:\\
Suppose, for simplicity,  that the classical action is given as $S=$Tr$P(D^2)$ where $P(x)$ is a polynomial that has precisely one minimum
$x_0$ on the real line. Then the minimum of $S$ is given for $D=x_0 \id$. (In fact, a ground state that is invariant under all unitaries 
must be of this form.)   Now, as explained above, the measure 
\[ {\rm d}\mu \, =\,  {\rm d }\lambda_1 \cdots  {\rm d }\lambda_n \quad \prod\limits_{i<j} (\lambda_i^2 -\lambda_j^2)^2 \quad e^{-S}  \]  
is spectral invariant. But then the vacuum expectation value
\[ \langle \lambda_k^2 - \lambda_l^2 \rangle = \int\, {\rm d}\mu\,\, (\lambda_k^2 - \lambda_l^2) \]
won't vanish in general. That is to say, the unitaries can not be represented on the Hilbert space of the quantum theory
in such a way that the vacuum state would be invariant. (Otherwise the expectation values of any two eigenvalues of $D$
-- viewed as observables -- would be
equal.) As we shall see in the examples, measures like the one above appear rather naturally due to the curved geometry
of the moduli space.\\
Nevertheless, the loss of spectral invariance can obviously be traced back to properties of the chosen measure,
and one might well (and should) ask whether such a choice is justified. Indeed, one might as well take the requirement
of spectral invariance of the vacuum as a criterion to prefer some measures over others.\\
On the other hand, it should always be kept in mind that the underlying ``manifold'' is zero-dimensional and accordingly 
there's no time and hence no canonical action of the symplectic group. So there's no canonical measure
for the path integral, since for (finite-dimensional) quantum mechanical systems the measure is singled out
by its invariance under canonical transformations combined with the given classical limit of the system.
To be more precise, the absence of canonical transformations should therefore be viewed as ambiguity of the definition of the 
``classical'' action as one might always redefine
\[ {\rm d}\mu \, e^{-S} \,= \,  {\rm d}{\tilde \mu} \, e^{-{\tilde S}}. \]
In fact, writing the above measure in the form $ {\rm d}\lambda_1\cdots {\rm d}\lambda_n\, e^{-{\tilde S}}$, one sees 
that the so defined action ${\tilde S}$ does not have a unique minimum. The ground state is degenerate with a nontrivial
action of the unitaries via permutations of the various minima, thus leading to a spontaneous breaking of spectral invariance
in the quantized theory.\\
This then points to the most important question we plan to investigate in this project:
\begin{center}
\em Can one define a classical limit for such systems ?
\end{center}
There need not exist an (unique) answer to this question, but if there is one, then it can only be found by a detailed
analysis of many examples. In the complementary project explained in \cite{RH,CP} the mathematical structure
of a pertubative treatment of a measure ${\rm d}\lambda_1\cdots {\rm d}\lambda_n\, e^{-S}$ is analyzed
in great depth by exploiting the noncommutativity of the underlying manifold.\\
In this illustrative paper we shall mainly consider Gaussian measures $S\sim$Tr$D^2$ on the moduli spaces.
We should stress once again that one can not always choose a spectral invariant measure, as there  are sometimes
observables which can not be expressed in terms of the eigenvalues of $D$.\\  
For the two-point space the  only interesting observable is the distance $d$ of the points,
which in the simplest example, when $D$ has only one (independent) eigenvalue $\lambda$, is given by
\[ d= \frac{1}{\lambda}. \]
Thus, for the Gaussians measure its vacuum expectation value
\[ \langle d \rangle = \int\limits_0^\infty\,\, {\rm d} \lambda \,\, \frac{1}{\lambda}\,\, e^{-t\lambda^2} = \infty \]
diverges, as had to be expected, since this corresponds to the unique ``classical'' ground state $D=0$.
However, in other examples, where $D$ has more than one eigenvalue, we find finite vacuum expectation values
for $d$, pointing to some attractive force between the points caused by quantum effects. \\
Another way to achieve finite expectation values of $d$ is to couple the system to fermions:
Then as 
\[ \int\limits_0^\infty {\rm d} \lambda \,\,e^{\langle \psi, D\psi\rangle } \sim \lambda^2 \]
one has
\[ \langle d \rangle_F  \sim \int\limits_0^\infty {\rm d} \lambda \,\,\lambda\, e^{-t\lambda^2} < \infty. \]
As mentioned above this paper does, however, not aim at illustrating these (and other) effects, but to 
work out the problems one is facing when trying to quantize finite spectral triples. So let's start with the mathematics.

\newpage

\section{Review: Finite spectral triples}  \label{review}
We start with
a very short survey of the main facts on finite (discrete) spectral
triples, thereby fixing the notation for the following sections. A detailed version can be found in \cite{TKendlicheTripel,MPendlicheTripel} 
or in \cite{diss}.
\\
The main ingredients of a spectral tripel are an (involutive) algebra
$\CA$, a Hilbert space $\CH$ that carries a representation of $\CA$ and a
selfadjoint operator $D$ (Dirac operator) acting on $\CH$. We restrict
our considerations to complex algebras. In addition one has (in the case of real triples) an
antilinear operator $J$ and for even dimensions a grading
$\Gamma$ on $\CH$.
\\
In the following the structure of $\CA$, $\CH$ and $D$ will be analyzed by
using the axioms for spectral triples. For finite spectral triples
the algebra and the
Hilbert space are finite dimensional and we repeat here only the
relevant axioms for that case: 
\begin{itemize} 
\item $\CA$ is a $C^*$--algebra. 
\item The Hilbert space $\CH$ carries a ($*$--) representation $\pi$ of $\CA$. 
\item $D$ is a selfadjoint operator on $\CH$. 
\item $\Gamma$ is a symmetry (i.e. $\Gamma = \Gamma^* , \Gamma^2 =1$) that commutes with $\pi$ and anticommutes with $D$: 
\[ [\pi (a),\Gamma ]=0\qquad \forall a\in \CA, \qquad\qquad  \qquad\Gamma D + D \Gamma =0 .\]  
\item $J$ is an antiunitary operator on $\CH$, which fulfills $J^2=1$. 
\item The map $\pi^o (a) := J \pi (a^*) J^{-1}$ is a representation of the opposite algebra $\CA^o$ that commutes with $\pi$
\[ \pi^o(a)\pi^o(b) = \pi^o(ba),  \qquad\qquad [\pi (a) , \pi^o (b) ] =0 \qquad\qquad  \forall a,b \in \CA. \]
\item ``Order one condition'' : 
\[ \big[ [D,\pi (a) ] , \pi^o (b) \big] = 0 \qquad\qquad\qquad  \forall a,b \in \CA .\]
\item ``Orientability'' : \[ \exists c \in \CA \otimes \CA^o \qquad \qquad \mbox{with} \qquad \qquad  \pi (c) = \Gamma . \]  
The representation of $\CA \otimes \CA^o$ is defined by $\pi (a \otimes b) := \pi (a) \pi^o (b)$. 
\item ``Poincar\'{e}--duality'':\\
The Fredholm index of the operator $D$ defines a nondegenerate intersection form on $K_\bullet (A)  \times K_\bullet (A)$. 
\item The following relations hold: \[ [J,D]=0, \qquad\qquad\qquad [J,\Gamma ] =0. \]
\end{itemize} 

One can then characterize the solutions $\CA$, $\CH$ and $D$ of these conditions. First of all, each finite dimensional 
complex $C^*$--algebra is a direct sum of matrix algebras
\[
 \CA = \bigoplus_{i=1}^k M_{n_i} (\mathbb{C} ) 
\] 
To analyze the structure of the Hilbert space we decompose it in the
following way: Define $P_i$ as the projector on the i-th subalgebra of
$\CA$
\begin{equation*} 
P_i := 0_{n_1 \times n_1} \oplus \ldots \oplus {\bf 1}_{n_i \times n_i} \oplus 0 \oplus \ldots \oplus 0  
\end{equation*} 
and set
\begin{equation} \label{defHij} 
\CH_{ij} := \pi (P_i) \pi^o (P_j)\, \CH 
\end{equation} 
where $\pi$ und $\pi^o$ are the representations of $\CA$ and $A^o$
respectively. Up to unitary equivalence, the only irreducible representation of $M_n (\mathbb{C}
)$ is $\C^n$, so each $\CH_{ij}$ has the form 
\begin{equation} 
\CH_{ij} = \mathbb{C}^{n_i} \otimes \mathbb{C}^{r_{ij}} \otimes \mathbb{C}^{n_j}
\end{equation} 
where the representation $\pi$ acts on the left factor in the
tensor product and $\pi^o$ on the right.\\
From the orientability condition it then follows that the grading $\Gamma$ acts on the Hilbert subspaces $\CH_{ij}$ only as
$\pm id$. If we define this sign as $\gamma_{ij}$ and moreover $q_{ij}
:= \gamma_{ij} r_{ij}$ the full information about the Hilbert space $\CH$
and the grading $\Gamma$ is encoded in the matrix $q$. \\ 
The reality operator $J$ maps $\CH_{ij}$ to $\CH_{ji}$ and from its invertibility it then follows that the matrix 
$q$ is symmetric. This matrix turns out to coincide with the
intersection form in K--theory defined by $D$ (the axiom of
Poincar\'{e}--duality then requires $q$ to be invertible).\\
For our project, the only interesting part of a spectral triple is the Dirac operator $D$. The axioms are 
strong restrictions on $D$, which we shall describe in the remainder of this section. With the definitions
\bea \label{bloeckevonD2}
D_{ij,kl} \,\, : \,\,  \CH_{kl} \to  \CH_{ij} \eea 
i.e  $D_{ij,kl} := \pi (P_i) \circ \pi^o (P_j) \circ D \circ \pi (P_k) \circ \pi^o (P_l)$
and 
\[
a_i := a P_i = P_i a 
\]
one can prove the following relations  
\begin{thm} (Structure of $D$) \label{strukturvonD} 
\begin{itemize} 
\item $D_{ij,kl} = D_{kl,ij}^*$  
\item $D_{ij,kl} = 0\quad$ if $\quad\gamma_{ij} = \gamma_{kl}$ 
\item $[D_{ij,il} , \pi (a_i) ] = 0 \qquad\qquad \forall a_i $ 
\item $[D_{ij,kj}, \pi^o (b_j) ] = 0 \qquad\qquad \forall b_j $ 
\item $D_{ij,kl} = 0\quad$ if $\quad i\ne k\quad$ and $\quad j \ne l$ 
\item $D_{ij,ij} = 0$ 
\end{itemize}   
Moreover, using the freedom in the choice of basis in $\C^{r_{ij}}$ it can be shown, 
that one can always choose a basis of $\CH$ with the following properties: 
\begin{itemize} 
\item $\{ v_1 , \ldots , v_n \}$ is a basis for $\CH_{ij}$ 
\item $\{ w_1 , \ldots , w_n \}$ is a basis for $\CH_{ji}$ 
\item $J v_k = w_k\quad$ if $\quad i \ne j$ 
\item $J v_k = J (x \otimes y \otimes z) = \bar{z} \otimes \bar{y} \otimes \bar{x}\quad $  if $\quad i = j$ 
\end{itemize}  
(As the proofs of this fact in the literature are actually not completely compelling, we 
present a new proof in the appendix.)\\
Then $D$ has the additional properties 
\begin{itemize} 
\item $D_{ij,kl} = \overline{D_{ji,lk} } \, $ for $i\ne j$ and $k \ne l$  
\item $D_{ii,kl} = J \circ D_{ii,lk} \circ J$
\item $D_{ij,kk} = J \circ D_{ji,kk} \circ J$ 
\end{itemize} 
\end{thm} 
In conclusion, given the algebra $\CA$ and the intersection form $q$ the dimensions of the Hilbert spaces in the 
composition (\ref{defHij}) are fixed and the structure of all possible Dirac operators is given by theorem \ref{strukturvonD}.  
As mentioned in the introduction, this result does, however, not provide a complete classification of finite spectral triples 
{\em up to unitary equivalence}:\\
While the freedom of choice for the basis of $\CH$ has been used to bring $\pi(\CA)$, $ \Gamma$ and $J$ into a canonical form,
it has not been investigated, whether the remaining freedom -- i.e. the unitaries which commute with  $\pi(\CA)$, $ \Gamma$ and $J$ --
can be used to restrict the space of Dirac operators. We shall return to this question in section \ref{moduli}.

\newpage
\section{Dirac operators and distances in the 2--point space} \label{sectiondistances} 
Before we turn our attention to the quantization of finite spectral triples, we shall 
work out the most important observable for the two-point space $\CA=\C^2$, namely the distance of the two
points. Other important observables are the eigenvalues of the Dirac operator, of course. How to obtain their
analytic expression in terms of the matrix entries of $D$ is clear, however.\\
\\
For a general commutative spectral triple, with $\CA=C^\infty(M)$ 
the geodesic distance of two points in $M$ can be recovered
from the purely algebraic data using Connes' celebrated  distance formula 
\[
d(p,q) =\sup_{f \in C^\infty (M)} \{ | f(p) - f(q)| : \| [D,f] \| \le 1. \}
\] 
(The Hilbert space $\CH$ where $D$ acts on is the space of square integrable spinors on $M$ and the 
representation of a function $f \in C^\infty (M)$ is given by pointwise multiplication.) \\  
This formula remains valid in case the algebra $\CA$ is not
longer commutative, if one replaces the ``points'' by the pure states
of the algebra. ( This corresponds to viewing $f(p)$ as $\delta_p (f)$ in the commutative case.) 
In the general case we write for pure states $\phi , \psi$:  
\[ 
d(\phi, \psi) = \sup_{a \in A} \{ | \phi (a) - \psi (a) | : \| [D,a] \| \le 1 \} 
\]

\vspace*{0.5cm}
\noindent
In the following 
we compute the distance-formula for the 2--point space, $A= \C \oplus \C$. We hope that this example also illustrates the 
formalism for finite spectral triples given in the previous section. 
There it was shown that
spectral triples for this algebra are determined by (invertible) $2
\times 2$ matrices with integer entries. If one takes into account the
symmetry of $q$, then the most general matrix is 
\begin{eqnarray*}   q = \left(
  \begin{array}{cc} k & -l \\ -l & m \end{array} \right) \qquad\qquad k,l,m
\in \mathbb{N} 
\end{eqnarray*} 
Here $\CH$,$\pi$ and $D$ are given by
\begin{eqnarray*} 
\CH & = & \CH_{11} \oplus \CH_{12} \oplus \CH_{21} \oplus \CH_{22} \\ 
  & \cong & \mathbb{C}^k \oplus \mathbb{C}^l \oplus \mathbb{C}^l \oplus \mathbb{C}^m 
\end{eqnarray*} 

\begin{eqnarray*} 
\pi (x,y) & = & \left( \begin{array}{cccc} x {\bf 1}_k & 0 & 0 & 0 \\ 
                                           0 & x {\bf 1}_l & 0 & 0 \\ 
                                           0 & 0 & y {\bf 1}_l & 0 \\ 
                                           0 & 0 & 0 & y {\bf 1}_m    \end{array} \right)  
\end{eqnarray*} 
\\ 
\\
\begin{eqnarray*} 
D & = & \left( \begin{array}{cccc} 0         & D_{11,12} & D_{11,21} & 0         \\ 
                                   D_{12,11} & 0         & 0         & D_{12,22} \\ 
                                   D_{21,11} & 0         & 0         & D_{21,22} \\ 
                                   0         & D_{22,12} & D_{22,21} & 0              \end{array} \right)  
\end{eqnarray*} 
Because of the symmetries of $D$ (Thm. \ref{strukturvonD}) this can be simplified to 
\begin{eqnarray*} 
D_{11,12} = \bar{D}_{11,21} = D^*_{12,11} = D^T_{21,11} & =: & \quad M\,\, :\,\, \mathbb{C}^l \to \mathbb{C}^k \\ 
D_{12,22} = \bar{D}_{21,22} = D^*_{22,12} = D^T_{22,21} & =: & \quad N \,\, :\,\, \mathbb{C}^m \to \mathbb{C}^l 
\end{eqnarray*} 
Thus one gets 
\begin{eqnarray*} 
D & = & \left( \begin{array}{cccc} 0   & M   & \bar{M} & 0          \\ 
                                   M^* & 0   & 0       & N          \\ 
                                   M^T & 0   & 0       & \bar{N}    \\ 
                                   0   & N^* & N^T     & 0              \end{array} \right)  
\end{eqnarray*} 
and with this notation one then obtains by a straightforward calculation:
\begin{eqnarray*} 
\Rightarrow [D, \pi (x,y)] & = &  (x-y) \cdot \left( \begin{array}{cccc} 0    & 0    & -\bar{M} & 0   \\ 
                                                                        0    & 0    & 0        & -N  \\ 
                                                                        M^T  & 0    & 0        & 0   \\ 
                                                                        0    & N^*  & 0        & 0  \end{array} \right)  =: (x-y) R \\                            
\end{eqnarray*} 
The norm of this one form $ [D, \pi (x,y)]$ that we shall need for the distance is then conveniently calculated as 
\begin{eqnarray*} 
\| R \|^2 & = & \| R^* R \| = \left\| \left(  
                \begin{array}{cccc} \bar{M} M^T & 0     & 0           & 0           \\ 
                                     0           & N N^* & 0           & 0           \\ 
                                     0           & 0     & M^T \bar{M} & 0           \\ 
                                     0           & 0     & 0            & N^* N       
                \end{array}   \right) \right\|  \\ 
          &   &  \\
          & = & \max \{ \| \bar{M} M^T \| , \| N N^* \| , \| M^T \bar{M} \| , \| N^* N \| \} .
\end{eqnarray*} 
But $\| \bar{M} M^T \| = \| M M^* \|$ (conjugation of the entries does not change the norm) and in addition $\| M M^* \| = \| M^* M \|$. 
For nonquadratic matrices this is not obvious but nevertheless true: $MM^*$ and $M^* M$ are selfadjoint matrices so their norm equals the biggest eigenvalue. However the eigenvalues are equal, though they have a different degree of degeneracy, as one can see from the following consideration: If $\lambda$ is an eigenvalue of  $MM^*$ for the  eigenvector $v$, then $M^* v$ is an eigenvector of $M^* M$ for the same eigenvalue $\lambda$. Hence one gets 
\begin{equation*} 
\| R \|^2 = \max \{ \| M^* M \|, \| N^* N \| \} 
\end{equation*} 
and finally 
\begin{equation*} 
\| [D , \pi (x,y) ] \| = |x-y| \cdot \sqrt{ \max ( \| M^* M \|, \| N^* N \| ) }
\end{equation*} 
As $ |x-y|$ is nothing but $|\phi(x,y) -\psi(x,y)|$ for the two pure states on $\C^2$, one immediately obtains:
\begin{eqnarray*} 
d(p_1 , p_2 ) & = &  \sup \{ |x-y| : \| [D , \pi (x,y) ] \| \le 1 \} \\ 
             &   &  \\ 
             & = & \left[ \max \left( \sqrt{\| M^* M \|} ,  \sqrt{\| N^* N \| } \right) \right]^{-1}  
\end{eqnarray*} 

It is important to note that this distance can -- apart from very particular examples -- not be expressed in terms of the
eigenvalues of $D$. This is due to the fact that -- as a consequence of the order-one condition -- the one-forms 
$[D, \pi(x,y)]$ have much fewer nonvanishing entries than $D$. The following example illustrates this 
effect of ``isospectral two-point spaces''.  
\begin{example} \label{dreieins} 
{\rm For the intersection form}  
\begin{eqnarray*}
q = \left( \begin{array}{cc}  1 & -1 \\ -1 & 1 \end{array} \right) 
\end{eqnarray*}
\end{example} 
the most general Dirac operator is of the form: 
\begin{eqnarray*}
D = \left( \begin{array}{cccc} 0 & m & \bar{m} & 0 \\ \bar{m} & 0 & 0 & \mu \\ m & 0 & 0 & \bar{\mu} \\ 0 & \bar{\mu} & \mu & 0 \end{array} \right) \quad m , \mu \in \C .
\end{eqnarray*}
It turns out that one of the two complex numbers (say $m$) could be chosen real positive. For the distance one then has  
\[
d(1,2) = \frac{1}{\max \{ m , |\mu | \} } ,
\] 
whereas the eigenvalues of the Dirac operator are given by 
\bea 
\lambda_{\pm}^2 = m^2 + |\mu|^2 \pm |m^2 + \mu^2 |
\eea
and thus depend on the phase of $\mu$ respectively the relative phase of $\mu$ and $m$. The reader then easily verifies,
that $d(1,2)$ cannot be expressed as a functional of $\lambda_{\pm}^2$ as only functions of the combination $\lambda_+^2+\lambda_-^2$
would be independent of the phase of $\mu$.\\ 
Nevertheless  the eigenvalues do set certain bounds  on the magnitude of $d(1,2)$, e.g. one has the relation: 
\[  
\frac{\sqrt{2}}{|\lambda_+|} \le d(1,2) \le \frac{2}{|\lambda_+|}  
\] 
\newpage

\section{Moduli spaces} \label{moduli} 
As mentioned in the introduction, in order to define a quantization (or ``sum over states'') of spectral triples with the help of an integral 
\bea 
{\cal Z} \sim \int \CD e^{-S(D)} 
\eea
we first need to classify equivalent spectral triples for a given algebra $\CA$ and given intersection form $q$. Let us briefly
recall the definition.
\begin{defn} \label{defequivalence} 
Two spectral triples $(A,\CH ,\pi,D,\Gamma,J)$ and $(A, \CH',\pi',D',\Gamma',J')$ are said to be equivalent, if there is a unitary operator $U: \CH \to \CH'$ with the properties that the following diagram commutes for $F=\pi (a) ,J,\Gamma,D$: 
\[    
\xymatrix{ \CH \ar[r]^U \ar[d]_F & \CH' \ar[d]^{F'} \\ \CH \ar[r]^U & \CH' } 
\]  
\end{defn} 
Since in our framework the dimensions of the Hilbert spaces are given by the matrix $q$ we can 
restrict the considerations to the same Hilbert space $\CH=\CH'$. 
Moreover the canonical form of $J$ and $\Gamma$ is achieved by a suitable choice of basis in $\CH$. But then, there still exist
unitary maps on $\CH$ which commute with $\pi, J$ and $\Gamma$. 
These transformations characterize the equivalence classes of Dirac operators we are looking for. 
The basic structure of these classes is quite easily described: If $U$ commutes with $\Gamma$, it must be block diagonal:
\[
U \CH_{ij} = \CH_{ij} 
\]
and because it commutes with the representation $\pi$ (and also with the opposite $\pi^o$) we find 
\[ 
U_{ij} = {\bf 1}_{n_i} \otimes u_{ij} \otimes {\bf 1}_{n_j} \quad \quad U_{ij} := U|_{\CH_{ij}} 
\]  
where now $u_{ij}$ is a unitary map $u_{ij} : \C^{r_{ij}} \to \C^{r_{ij}}$. The relation $[U,J]=0$ then leads to the further restriction 
\[
u_{ij} = \overline{u_{ji}} 
\]  
In particular, the matrices $u_{ii}$ are orthogonal matrices (with real entries).
Thus our task consists of finding equivalence classes for the relation 
\[
D \sim U D U^* , \qquad \qquad \qquad D_{ij,kl} \sim u_{ij} D_{ij,kl} u_{kl}^* 
\]
where $U$ is restricted by the explained relations. \\ 
We start the business with simple examples which demonstrate the problem, 
the more general case is discussed in a forthcoming paper. 
The algebra in the following examples is always chosen to be $A = \C \oplus \C$.  

\begin{example} \label{daseinfachste}  
\[ 
q = \left( \begin{array}{cc}  1 & -1 \\ -1 & 0   \end{array} \right) 
\] 
\end{example} 
The corresponding Hilbert space is $ \CH = \CH_{11} \oplus \CH_{12} \oplus \CH_{21} \cong \C^3$ and the Dirac operator has the form 
\[ 
D = \left( \begin{array}{ccc}  0 & \bar{m} & m \\ m & 0 & 0 \\ \bar{m} & 0 & 0   \end{array} \right) \quad m \in \C.
\]
Now consider a unitary matrix $U$ obeying the discussed restrictions. It has the form 
\[ 
U = \left( \begin{array}{ccc}  1 & 0 & 0 \\ 0 & a & 0 \\ 0 & 0 & \bar{a}   \end{array} \right) \quad a \in U(1) 
\]
and so we have 
\[
U D U^* = \left( \begin{array}{ccc}  0 & \bar{m} \bar{a} & ma \\ m a  & 0 & 0 \\ \bar{m} \bar{a}  & 0 & 0   \end{array} \right) .
\]
Obviously one can achieve with an appropriate choice of $a$  ($a= \frac{\bar{m}}{|m|}$) that 
\bea \label{Dfuerseinfachste} 
D =  \left( \begin{array}{ccc}  0 & m & m \\ m & 0 & 0 \\ m & 0 & 0   \end{array} \right) \quad m \in \R , m \ge 0 .
\eea 

In the general case the intersection form is  
\[ 
q =\left( \begin{array}{cc} k & -l \\ -l &
    m \end{array} \right) \quad k,l,m \in \mathbb{N} 
\]
and the corresponding (general) Dirac operator 

\begin{eqnarray*} 
D & = & \left( \begin{array}{cccc} 0   & M   & \bar{M} & 0          \\ 
                                   M^* & 0   & 0       & N          \\ 
                                   M^T & 0   & 0       & \bar{N}    \\ 
                                   0   & N^* & N^T     & 0              \end{array} \right)  
\end{eqnarray*} 
where
\begin{eqnarray*} 
M \,\, & : & \,\,\mathbb{C}^l \to \mathbb{C}^k \\ 
N \,\, & : & \,\, \mathbb{C}^m \to \mathbb{C}^l, 
\end{eqnarray*} 
as has already been mentioned in the previous section. The unitary transformations are given by  
\begin{eqnarray*} 
U = \left( \begin{array}{cccc}     A  & 0  & 0 & 0   \\ 
                                   0  & V  & 0 & 0   \\ 
                                   0  & 0  & \bar{V} & 0   \\ 
                                   0  & 0  & 0 & B    \end{array} \right)  \quad A \in O(k) ,\quad V \in U(l) , \quad B \in O(m) 
\end{eqnarray*} 
and for the transformed Dirac operator one gets 
\[
U D U^* = \left( \begin{array}{cccc}  
0 & AMV^* & A \bar{M} V^T & 0 \\ 
VM^* A^T & 0 & 0 & VNB^T \\ 
\bar{V} M^T A^T & 0 & 0 & \bar{V} \bar{N} B^T \\ 
0 & BN^*V^* & BN^TV^T& 0 
\end{array}  \right) .
\] 
There are of course only two independent blocks $AMV^*$ and $VNB^T$, because
of the relations 
\[
\overline{AMV^*} = A \bar{M} V^T , \qquad\qquad \overline{VNB^T} = \bar{V} \bar{N} B^T 
\]
and the fact that $UDU^*$
is selfadjoint. Our task is to find 
canonical forms for the matrices $M$ and $N$ under these transformations. A general solution for this problem is still in work. 
So let's have a look at further examples: 
\begin{example} \label{beisp4.2}
\[
q = \left( \begin{array}{cc}  1 & -1 \\ -1 & 1 \end{array} \right) 
\] 
\end{example} 
(We do not worry about the fact that $q$ is not invertible.) Here the Hilbert space is $\C^4$ and 
\[
D = \left( \begin{array}{cccc} 0 & m & \bar{m} & 0 \\ \bar{m} & 0 & 0 & \mu \\ m & 0 & 0 & \bar{\mu} \\ 
0 & \bar{\mu} & \mu & 0 \end{array} \right) \quad m , \mu \in \C
\] 
The admissible unitary transformations are 
\[ 
U = \left( \begin{array}{cccc}  
1 & 0 & 0 & 0 \\ 
0 & a & 0 & 0 \\ 
0 & 0 & \bar{a} & 0 \\ 
0 & 0 & 0 & 1  \end{array} \right) \quad a \in U(1) 
\]
so only one complex phase can be eliminated in 
\[ 
U D U^* = \left( \begin{array}{cccc} 
0         & m \bar{a}         & \bar{m} a & 0                 \\ 
\bar{m} a & 0                 & 0         & \mu a             \\ 
m \bar{a} & 0                 & 0         & \bar{\mu} \bar{a} \\ 
0         & \bar{\mu} \bar{a} &\mu a      & 0 
\end{array} \right) 
\] 
by choosing for example $a = \frac{m}{|m|}$. Hence, in this case the equivalence classes of 
Dirac operators can be parametrized by a real nonnegative $m$ and a complex number $\mu$. 
\begin{example} \label{Beispiel4.3} 
\[ 
q = \left( \begin{array}{cc}  2 & -2 \\ -2 & 0   \end{array} \right) .
\]
\end{example} 
Here $\CH = \C^6$ and a Dirac operator looks like
\[
D = \left( \begin{array}{ccc} 0 & m & \bar{m} \\ m^* & 0 & 0 \\ m^T & 0 & 0 \end{array} \right) \quad m \in \C^{2 \times 2} .
\] 
Unitary equivalence of the corresponding spectral triples is given by transformations of the form 
\[ 
U = \left( \begin{array}{ccc} A & 0 & 0 \\ 0 & V & 0 \\ 0 & 0 & \bar{V} \end{array} \right) \quad A \in O(2), V \in U(2) .
\] 
This leads to 
\[ 
UDU^*  = \left( \begin{array}{ccc} 0 & AmV^* & A\bar{m} V^T \\ Vm^*A^T & 0 & 0 \\ \bar{V} m^T A^T & 0 & 0 \end{array} \right) 
\] 
and so we have to look for a representative in the class of Dirac operators under the transformation 
\[ 
m \to AmV \quad A \in O(2), V \in U(2) 
\] 
The solution of this problem is given by the following 
\begin{thm} \label{OCU} 
Let $m$ be an arbitrary nonsingular $2\times 2$-matrix. Then there is a unique positive (selfadjoint) matrix $C$ of the following form 
\[ 
C = \left( \begin{array}{cc} a & ic \\ -ic & b \end{array} \right) \quad a,b,c \in \R , \quad a \ge b \ge 0 , ab \ge c^2 
\] 
as well as a unitary matrix $U$ and an orthogonal matrix $O$ (both unique if $a \ne b$) such that 
\[ 
m = OCU 
\] 
\end{thm} 
PROOF: The technical basis for the proof is the following quick calculation: 
\[ 
\left( \begin{array}{cc} \cos \alpha & \sin \alpha \\ - \sin \alpha & \cos \alpha \end{array} \right) \left( \begin{array}{cc} x & z \\ 
\bar{z} & y \end{array} \right) \left( \begin{array}{cc} \cos \alpha & - \sin \alpha \\ \sin \alpha & \cos \alpha \end{array} \right)
 = \left( \begin{array}{cc}  X & Z \\ \bar{Z} & Y \end{array} \right) 
\] 
with 
\bea 
 X & = & x \cos^2 \alpha + y \sin^2 \alpha + 2 (\mbox{Re} \,  z) \sin \alpha \cos \alpha  \\ 
 Z & = & (y-x) \sin \alpha \cos \alpha + (\mbox{Re} \, z) (\cos^2 \alpha - \sin^2 \alpha ) + i  (\mbox{Im} \, z)  \label{ReT12} \\ 
 Y & = & x \sin^2 \alpha + y \cos^2 \alpha - 2  (\mbox{Re} \, z) \sin \alpha \cos \alpha  \label{ocubedingungen}    
\eea
In particular the imaginary part of the off diagonal entry is invariant under orthogonal transformations. 
We first show the existence of the claimed decomposition and then uniqueness. \\ 
{\em Existence}: \\
Each nonsingular matrix $m$ can be written in spectral decomposition as 
\[ 
m = T U 
\] 
where $T$ is positive (selfadjoint). Therefore we only need to show that there exists an orthogonal matrix $O$ and $C$ of the claimed form 
with $T=OCO^T$ (note that $O^T U$ is unitary). Alternatively it is sufficient to show the existence of an orthogonal matrix 
that cancels the real part of the off diagonal element in $T$, such that $T$ is transformed into the 
desired form $C$. With (\ref{ReT12}) this leads to the equation 
\[ 
(y-x) \sin \alpha \cos \alpha + (\mbox{Re} \, z) (\cos^2 \alpha - \sin^2 \alpha ) = 0 
\]
and after elementary transformations we get 
\[ 
\frac{y-x}{\mbox{Re} \, z} = - 2 \cot (2\alpha ) 
\]
which has a solution $\alpha$ for all possible values of $x,y,z$.  This then proves the existence of the decomposition. 
\\
{\em Uniqueness of the matrices $C,U$ and $O$}:\\ 
Suppose there are two possibilities for the decomposition 
\[ 
m = O_1 C_1 U_1 =   O_2 C_2 U_2 . \]
Then it follows
\[
C_2 =  O_2^T O_1 C_1 U_1 U_2^* 
\] 
and since $O^T U$ is unitary one can also consider 
\beas 
 C_2  = O C_1 O^T U  \iff C_2 U^* = OC_1 O^T  
\eeas 
The determinants of $C_1$ and $C_2$ are positive (real) implying $U\in SU(2)$. Moreover (because $OC_1 O^T$ is selfadjoint) we must have 
 \beas 
 U C_2  \stackrel{!}{=} C_2 U^* .
\eeas 
If we parametrize 
\beas 
 U = \left( \begin{array}{cc} \alpha & \beta \\ - \bar{\beta} & \bar{\alpha} \end{array} \right)
\eeas 
and $C_2$ as above we get the conditions 
\beas
a_2 \alpha -i c_2 \beta & \stackrel{!}{=} & a_2 \bar{\alpha} + i c_2 \bar{\beta}  \\ 
a_2 \beta & \stackrel{!}{=} & -b_2 \beta \\  
-i \bar{\beta} c_2 + \bar{\alpha} b_2  & \stackrel{!}{=} & i c_2 \beta + \alpha b_2  
\eeas 
Because $C_2$ is assumed to be positive we have $a_2 ,b_2 \ge 0$ and so the second equation gives two cases: $a_2 = b_2 =0$ or $\beta =0$. \\ 
In the first case, since $a_2b_2 \ge c_2^2$, we would have $C_2=C_1=0$. This implies $m=0$, which must be excluded. 
The case $\beta =0$ implies $\alpha =1$ (unitarity condition and third equation) which gives $U={\bf 1}$ and we only have to show 
$O= {\bf 1}$. But now (\ref{ReT12}) shows that $c_2 = c_1$ in any case (the imaginary part of the offdiagonal element is unaffected 
by orthogonal transformations). But then, because $C_1$ and $C_2$ have the same trace and the same determinant we know that either 
$a_1 = a_2, b_1 = b_2$ or $a_1 = b_2, a_2 = b_1$. If $a_1 \ne b_1$ the condition $a_1 \ge b_1$ rules out one possibility and the 
decomposition is unique. If $a_1 = b_1$ the decomposition is unique up to a $\pi/2$--rotation, since 
\beas 
\left( \begin{array}{cc} 0 & 1 \\ -1 & 0 \end{array} \right) \left( \begin{array}{cc} a & ic \\ -ic & a \end{array} \right) \left( \begin{array}{cc} 0 & -1 \\ 1 & 0 \end{array} \right) = \left( \begin{array}{cc} a & ic \\ -ic & a \end{array} \right) 
\eeas 
This shows the claimed uniqueness of the decomposition and completes the proof. \hfill \checkmark  

\begin{example} \label{2-1-10} 
\beas 
q = \left( \begin{array}{cc}  2 & -1 \\ -1 & 0 \end{array} \right) 
\eeas 
\end{example} 
The most general Dirac operator is 
\beas 
D = \left( \begin{array}{cccc} 0 & 0 & m_1 & \bar{m_1} \\ 
                               0 & 0 & m_2 & \bar{m_2} \\ 
                               \bar{m_1} & \bar{m_2} & 0 & 0  \\ 
                               m_1 & m_2 & 0 & 0  \end{array} \right) \qquad \qquad m_1 , m_2 \in \C 
\eeas
acting on $\CH = \CH_{11} \oplus \CH_{12} \oplus \CH_{21} \cong \C^2 \oplus \C \oplus \C$ and admissible unitary transformations are 
\beas 
U  = \left( \begin{array}{cccc} R &  0 & 0 \\ 
                               0 & e^{i \varphi} & 0   \\ 
                               0 & 0 & e^{-i \varphi}  \end{array} \right) \qquad\qquad R \in O(2),\quad  e^{i \varphi} \in U(1) 
\eeas 
If we put $\vec{m} := \left( \begin{array}{c} m_1 \\ m_2  \end{array} \right)$, then the Dirac operator can be written as 
\beas 
D = \left( \begin{array}{ccc} 0 & \vec{m} & \vec{\bar{m}} \\ \vec{m}^* & 0 & 0 \\ \vec{m}^T & 0 & 0     \end{array} \right) 
\eeas 
and unitary transformations lead to 
\beas 
UDU^* & = & \left( \begin{array}{ccc} R &  0 & 0 \\ 
                               0 & e^{i \varphi} & 0   \\ 
                               0 & 0 & e^{-i \varphi}     \end{array} \right) 
\left( \begin{array}{ccc} 0 & \vec{m} & \vec{\bar{m}} \\ \vec{m}^* & 0 & 0 \\ \vec{m}^T & 0 & 0     \end{array} \right) 
\left( \begin{array}{ccc}  R^T &  0 & 0 \\ 
                               0 & e^{-i \varphi} & 0   \\ 
                               0 & 0 & e^{i \varphi}    \end{array} \right) \nonumber \\ 
 & = & \left( \begin{array}{ccc}  0 & R \vec{m} e^{-i\varphi} & R \vec{\bar{m}} e^{i \varphi} \\ e^{i\varphi} \vec{m}^* R^T & 0 & 0 \\ 
e^{-i\varphi} \vec{m}^T R^T & 0 & 0 \end{array} \right)
\eeas

Again one can find representatives of the equivalence classes using the following 
\begin{thm} \label{RmUtheorem} 
Let $\vec{m} = \left( \begin{array}{c} m_1 \\ m_2 \end{array} \right) \in \C^2$. Then there exist $R \in O(2), \varphi \in [0,2\pi[$ and 
a unique $\psi \in [0,\frac{\pi}{2} ]$ such that 
\bea \label{RmU} 
 R \vec{m} e^{i\varphi}  = \frac{|\vec{m} |}{\sqrt{2}} \left( \begin{array}{c} 1 \\  e^{i\psi} \end{array} \right)  
\eea
So Dirac operators are parametrized by vectors of the form $ \frac{\rho}{\sqrt{2}} \left( \begin{array}{c} 1 \\  e^{i\psi} \end{array} \right) , \psi \in [0,\frac{\pi}{2} ]$, i.e. a ``quarter of a cone'' in $\C^2$. 
\end{thm} 
PROOF: The proof is similar to theorem \ref{OCU}, so readers who
are not interested in technical details can easily skip it. We first
show existence of the decomposition:\\ 
For $\left( \begin{array}{c} m_1 \\ m_2 \end{array} \right)$ exists $R \in SO(2)$ with the property 
\bea \label{gleicherbetrag} 
R \left( \begin{array}{c} m_1 \\ m_2 \end{array} \right) = \left( \begin{array}{c} u \\ v \end{array} \right) 
\quad \mbox{where} \quad |u| = |v| 
\eea
To prove (\ref{gleicherbetrag}) we put 
\beas
\vec{m} = \left( \begin{array}{c} r_1 e^{i\theta_1} \\ r_2 e^{i\theta_2} \end{array} \right)
\eeas
and calculate 
\beas 
R \vec{m} 
 & = &  \left( \begin{array}{c} r_1 \cos \rho \, e^{i\theta_1} - r_2 \sin \rho \, e^{i\theta_2} \\ r_1 \sin \rho \, e^{i\theta_1} + r_2 
\cos \rho \, e^{i\theta_2} \end{array} \right)
\eeas 
Now we get for the (square of the) modulus of the two entries 
\beas 
| r_1 \cos \rho \, e^{i\theta_1} - r_2 \sin \rho \, e^{i\theta_2} |^2 & = & |r_1 \cos \rho  - r_2 \sin \rho \, e^{i(\theta_2 - \theta_1 )} |^2   \\ 
 & = & | r_1 \cos \rho - r_2 \sin \rho \cos \Delta - i r_2 \sin \rho \sin \Delta |^2 \quad ( \Delta := \theta_2 - \theta_1 ) \\ 
 & = & (r_1 \cos \rho - r_2 \sin \rho \cos \Delta )^2 + (r_2 \sin \rho \sin \Delta )^2 \\ 
 & = & r_1^2 \cos^2 \rho + r_2^2 \sin^2 \rho - 2 r_1 r_2 \sin \rho \cos \rho \cos \Delta \\ 
| r_1 \sin \rho \, e^{i\theta_1} + r_2 \cos \rho \, e^{i\theta_2} |^2 
 & = & r_1^2 \sin^2 \rho + r_2^2 \cos^2 \rho + 2r_1r_2 \sin \rho \cos \rho \cos \Delta 
\eeas 
Equality of the two terms leads to 
\beas
 (r_1^2 - r_2^2 )(\cos^2 \rho - \sin^2 \rho ) =  4r_1 r_2 \sin \rho \cos \rho \cos \Delta \nonumber \eeas
\bea
\underbrace{\frac{\cos^2 \rho - \sin^2 \rho }{\sin \rho \cos \rho }  }_{= 2 \cot 2\rho } 
& = & \frac{4r_1 r_2 \cos \Delta }{r_1^2 - r_2^2 } \label{Uniq12u} 
\eea
which has always a solution $\rho$. Thus there is $R\in SO(2)$ and $\theta_1 , \theta_2 \in [0,2\pi[$ with the property 
\beas 
R \vec{m} = r \left( \begin{array}{c} e^{i \theta_1} \\  e^{i \theta_2} \end{array} \right)  
\eeas 
The following matrices belong to $O(2)$ 
\beas 
 \left( \begin{array}{cc} 1 & 0 \\ 0 & -1  \end{array} \right) , \left( \begin{array}{cc} -1 & 0 \\ 0 & 1  \end{array} \right) ,  
\left( \begin{array}{cc} -1 & 0 \\ 0 & -1  \end{array} \right) , \left( \begin{array}{cc} 0 & 1 \\ 1 & 0  \end{array} \right)   
\eeas 
so one can always achieve the case $0 \le \theta_1 \le \theta_2 < \pi$. For example 
\beas 
\left( \begin{array}{cc} 1 & 0 \\ 0 & -1  \end{array} \right) \left( \begin{array}{c} e^{i \theta_1} \\  e^{i \theta_2} \end{array} \right) = 
\left( \begin{array}{c} e^{i \theta_1} \\  e^{i (\theta_2 - \pi) } \end{array} \right)
\eeas 
It only remains to show that one can choose $\psi \in [0, \frac{\pi}{2} ]$ in the claimed decomposition. To that end consider 
\beas 
 R \vec{m} e^{i\varphi}  =  \frac{|\vec{m} |}{\sqrt{2}} \left( \begin{array}{c} 1 \\  e^{i\psi} \end{array} \right) \quad \psi 
\in [\frac{\pi}{2} , \pi [ \eeas
\beas 
 \Rightarrow \left( \begin{array}{cc} 0 & -1 \\ 1 & 0  \end{array} \right) R \vec{m} e^{i\varphi} e^{i(\pi - \psi)} =   
\frac{|\vec{m} |}{\sqrt{2}} \left( \begin{array}{c} 1 \\  e^{i(\pi - \psi)} \end{array} \right)   
\eeas
and  $\psi' := \pi - \psi$ gives the desired result $\psi' \in [0, \frac{\pi}{2} ]$. 
\\
Now we examine the uniqueness of the decomposition: Suppose there are
$R_1 , \varphi_1 , \psi_1$ and $R_2 , \varphi_2 , \psi_2$ such that
\bea \label{Uniq1} R_1 \vec{m} e^{i \varphi_1} & = & \frac{|\vec{m}
  |}{\sqrt{2} } \left( \begin{array}{c} 1 \\ e^{i \psi_1 } \end{array}
\right) 
\eea 
\bea \label{Uniq2} R_2 \vec{m} e^{i \varphi_2} & = &
\frac{|\vec{m} |}{\sqrt{2} } \left( \begin{array}{c} 1 \\ e^{i \psi_2
      } \end{array} \right) 
\eea 
Now equation (\ref{Uniq1}) implies
\bea \label{Uniq3} \vec{m} & = & \frac{|\vec{m} |}{\sqrt{2} } R_1^T
\left( \begin{array}{c} 1 \\ e^{i \psi_1 } \end{array} \right)
e^{-i\varphi_1} \eea (\ref{Uniq3}) in (\ref{Uniq2}) leads to \bea R_2
R_1^T \left( \begin{array}{c} 1 \\ e^{i \psi_1 } \end{array} \right)
e^{i(\varphi_2 - \varphi_1)} & = & \left( \begin{array}{c} 1 \\ e^{i
      \psi_2 } \end{array} \right) \eea 
Therefore it is sufficient to
consider the following equation: \bea \label{Uniq6} R \left(
  \begin{array}{c} 1 \\ e^{i \psi_1 } \end{array} \right) e^{i
  \varphi} & = & \left( \begin{array}{c} 1 \\ e^{i \psi_2 }
  \end{array} \right) 
\eea
At this point it is necessary to consider different cases. \\
{\em Case 1}:\\ 
$R \in SO(2)$ 
\bea \label{Uniq7} \Rightarrow R & =
& \left( \begin{array}{cc} \cos \rho & \sin \rho \\ - \sin \rho & \cos
    \rho \end{array} \right) \eea 
(\ref{Uniq7}) in (\ref{Uniq6}) leads to 
\beas \left( \begin{array}{cc} \cos \rho & \sin \rho \\ - \sin \rho
    & \cos \rho \end{array} \right) \left( \begin{array}{c} e^{i
      \varphi} \\ e^{i(\varphi + \psi_1)} \end{array} \right) =  
\left( \begin{array}{c} 1 \\ e^{i\psi_2} \end{array} \right) \eeas 
and we get the two following equations: 
\bea
\cos \rho \, e^{i \varphi} + \sin \rho \, e^{i(\varphi + \psi_1)} & = & 1 \label{Uniq8} \\
- \sin \rho \, e^{i \varphi} + \cos \rho \, e^{i(\varphi + \psi_1)} &
= & e^{i \psi_2} \label{Uniq9} 
\eea 
The modulus of the left
hand side of (\ref{Uniq8}) is calculated as: 
\beas
| \cos \rho \, e^{i \varphi} + \sin
\rho \, e^{i(\varphi + \psi_1)} |^2 =  1 + 2 \sin \rho \cos \rho
\cos \psi_1 
\eeas 
and so (\ref{Uniq8}) gives the condition 
\bea
\label{Uniq10} \sin \rho \cos \rho \cos \psi_1 & = & 0 
\eea
(We could have used equation (\ref{Uniq9}) instead, which gives the same result.) Now we have to treat the three different cases for 
which the product can be zero: \\
{\em Case 1.1}: $\sin \rho = 0$.\\ 
This implies 
\beas
R = \left( \begin{array}{cc} 1 & 0 \\  0 & 1 \end{array} \right) & \Rightarrow & \varphi = 0 , \, \psi_1 = \psi_2 \\
\mbox{or} \quad R = \left( \begin{array}{cc} -1 & 0 \\ 0 & -1
  \end{array} \right) & \Rightarrow & e^{i \varphi} = -1 , \, \psi_1 =
\psi_2 
\eeas
\noindent 
{\em Case 1.2}:$\cos \rho = 0 $.  
\beas \Rightarrow R = \left(
  \begin{array}{cc} 0 & 1 \\ -1 & 0 \end{array} \right) \quad
\mbox{or} \quad R = \left( \begin{array}{cc} 0 & -1 \\ 1 & 0
  \end{array} \right) \eeas 
Inserting $R = \left( \begin{array}{cc} 0 &
    1 \\ -1 & 0 \end{array} \right)$ into (\ref{Uniq6}) gives \beas
\left( \begin{array}{c} e^{i(\varphi + \psi_1)} \\ - e^{i \varphi}
  \end{array} \right) = \left( \begin{array}{c} 1 \\ e^{i \psi_2}
  \end{array} \right) \eeas and this leads to $\varphi = 2\pi - \psi_1$
and $\psi_1 + \psi_2 = \pi$ so only $\psi_1 = \psi_2 = \frac{\pi}{2}$
is a solution in the interval $[0,\frac{\pi}{2}]$.
The case $R = \left( \begin{array}{cc} 0 & -1 \\  1 & 0 \end{array} \right)$ is analogous. \\
{\em Case 1.3}: $\cos \psi_1 = 0 \Rightarrow e^{i \psi_1} = i$.\\
This leads to the equation \beas \left( \begin{array}{cc} \cos \rho &
    \sin \rho \\ - \sin \rho & \cos \rho \end{array} \right) \left(
  \begin{array}{c} 1 \\ i \end{array} \right) e^{ - i \rho} = \left(
  \begin{array}{c} 1 \\ i \end{array} \right) \eeas
which is fulfilled for all values of $\rho$ but gives $\psi_1 = \psi_2$.  \\
{\em Case 2}: $R \in O(2) , R \notin SO(2)$.\\
This case can be treated in
the same way and leads to the same results, so the proof is complete.
\hfill \checkmark

\newpage

\section{Quantization} \label{sectionquantization} 
Finally, after collecting all the necessary prerequisites, we come to our main objective,
the path integral quantization of finite spectral triples.
\\
Path integrals are based on the idea that each possible
state, which the system can pass on its way from the initial to the
final state, contributes to the transition amplitude. In the context
of gravity one would have to find a way to sum over Lorentzian manifolds, a problem that is not yet solved in $(3+1)$ dimensions. In
our framework of finite geometries, however, it is straightforward to define such a
summation in certain examples, once the moduli space is found. 
\\
For a given spectral triple $(A,q)$ we want to define a state sum 
\beas
{\cal Z} = {\cal N} \int \CD D \, e^{-S(D)} 
\eeas 
where the curly $D$ denotes the invariant measure that we are hunting for.
A ``classically'' (stable) spectral invariant action $S(D)$ on finite dimensional Hilbert spaces
can be written in the form 
\beas S(D) = \sum_{k=-\infty}^\infty t_k
\mbox{tr} D^{2k} 
\eeas 
It is sufficient to sum over even exponents since $\mbox{tr} D^{2k+1}
=0$ because $D$ anticommutes with the grading $\Gamma$.  
As we stressed in the introduction, one should not take the term ``classical action''
too seriously, as the absence of time and thus of canonical transformations makes the definition
of a classical limit of the system rather difficult.
\\
We shall sometimes also
couple the system to fermions $\psi\in \CH$, whose classical action is then given as 
\[S_{\rm ferm} = \langle \psi |\,D\, |\psi\rangle. \] 
We call them fermions as we quantize them according to Fermi-Dirac-statistics, though there is of course no
spin-statistics theorem that would tell us to do so. In particular, if we integrate the
fermions out, as we shall do throughout this paper, the effective action is given as
\beas 
{\cal Z}_F = {\cal N}' \int {\CD} D \, e^{-S_F(D)} 
\eeas 
where 
\[ S_F(D) = S(D)-\ln{\rm det}D .\]
Note that in most cases $\det D$ vanishes identically, so for a sensible definition of the fermionic
action we must calculate $\det D$ on the complement of the kernel of
$D$.
The normalization constants ${\cal N}$ and ${\cal N}'$ will sometimes be chosen in such a way that ${\cal Z} = 1$ and  
${\cal Z}_F =1$ for the gaussian $S= \frac{t}{2} \mbox{tr} D^2$, but in this paper we shall choose them such as to normalize
the vacuum to unity, i.e. $\langle 0| 0\rangle = {\cal Z} = 1$. \\ \\ 
Let's now come back to concrete examples. Again we only treat the algebra $\C \oplus \C$. 
\begin{example} 
\rm Consider the intersection form 
\beas 
q = \left( \begin{array}{cc} 1 & -1 \\ -1 & 0 \end{array} \right) 
\eeas 
\rm which is discussed in example \ref{daseinfachste}. 
\end{example} 
Each class of equivalent Dirac operators has a representative of the form 
\beas 
D =  \left( \begin{array}{ccc}  0 & m & m \\ m & 0 & 0 \\ m & 0 & 0   \end{array} \right) \quad m \in \R , m \ge 0 
\eeas 
as has been shown in equation (\ref{Dfuerseinfachste}). 
$D$ has eigenvalues $0,\pm m \sqrt{2}$ and the kernel of $D$ is spanned by a fixed vector $(0,1,-1)^T$. 
Neglecting the kernel one gets for the determinant 
\beas 
\det D = -2 m^2 .
\eeas
Since $D$ has only one independent eigenvalue $\lambda = m \sqrt{2}$ the invariant measure is clear: 
\beas
{\cal Z} = {\cal N} \int\limits_0^\infty {\rm d} \lambda \, e^{-S(\lambda)} 
\eeas
respectively (remember $\det D = -2m^2 = - \lambda^2$)  
\beas
{\cal Z}_F = - {\cal N}' \int\limits_0^\infty  {\rm d} \lambda \,  \lambda^2 \, e^{-S(\lambda)} .
\eeas
Fixing the constants $\cal N$ and $\cal N'$ we get 
\beas 
{\cal Z} &  = &  2\sqrt{\frac{t}{\pi}} \int\limits_0^\infty \, {\rm d} \lambda\,\, e^{-t \lambda^2} \\ 
{\cal Z}_F &  = &  4\sqrt{\frac{t^3}{\pi}} \int\limits_0^\infty  \, {\rm d} \lambda\,\,\lambda^2 e^{-t \lambda^2} 
\eeas

In section \ref{sectiondistances} we calculated for the distance $d(1,2) =
\frac{1}{m} = \frac{\sqrt{2}}{\lambda}$, so for the vacuum expectation values
we end up with the following expressions: 
\beas 
\langle d(1,2) \rangle & = &  4\sqrt{\frac{t}{2\pi}} \int\limits_0^\infty \, {\rm d} \lambda\,\,  \frac{1}{\lambda} e^{- S(\lambda)} 
\eeas 
\beas 
\langle d(1,2) \rangle_F & = &  8\sqrt{\frac{t^3}{2 \pi}} \int\limits_0^\infty \, {\rm d} \lambda\,\, \lambda \, e^{- S(\lambda)}.
\eeas 
That leads us to a first interesting observation:
\begin{lema} \label{lemmaexpdist} 
For each polynomial action $S(D) = \sum\limits_{k=0}^n t_k \lambda^{2k}$ the vacuum expectation value of the distance is 
\beas 
\langle d(1,2) \rangle  = \infty 
\eeas 

If we have in addition $t_{-1}\neq0$, we get $ \langle d(1,2)\rangle < \infty$. 
On the other hand this value remains always finite in the fermionic case: 
\beas
 \langle d(1,2)\rangle_F < \infty
\eeas 
For $S = t \lambda^2$ we obtain 

\beas 
\langle d(1,2)\rangle_F =4\sqrt{\frac{t}{2\pi}}
\eeas 
\end{lema}   
The proof is obvious and we skip it here. This example shows that terms of the form $t_{-1} \lambda^{-2}$ can be used to 
regularize the vacuum expectation value of the distance. The same effect is due to the coupling of fermions: 
It leads to an attractive force between the points, which is strong enough to give finite results. \\ \\ 
After this brief warm-up, we now come to more sophisticated examples.
In the next example we explicitly construct the invariant measure on the space of Dirac operators by using the 
results from the previous section and the Faddeev--Popov method. 

\begin{example} \label{bspFP} 
\beas 
q= \left( \begin{array}{cc} 2 & -2 \\ -2 & 0 \end{array} \right) 
\eeas 
\end{example} 
Dirac operators and representatives of the equivalence classes are worked out in example \ref{Beispiel4.3}: A general $D$ is of the
form 
\beas 
D = \left( \begin{array}{ccc} 0 & m & \bar{m} \\ m^* & 0 & 0 \\ m^T & 0 & 0 \end{array} \right) \qquad\qquad  m \in \C^{2 \times 2} 
\eeas 
and representatives are given by 
\bea \label{DundC} 
 D = \left( \begin{array}{ccc} 0 & C & \bar{C} \\ C  & 0 & 0 \\  \bar{C} & 0 & 0 \end{array} \right) 
\qquad \mbox{where} \quad C = \left( \begin{array}{cc} a & ic \\ -ic  & b  \end{array} \right) \qquad a \ge b \ge 0 , ab \ge c^2 
\eea
The construction of the invariant measure can be carried out as follows: \\ 
We know that the equivalence classes are given by $ m \sim OmU$ , with $O \in O(2)$  and $U \in U(2)$. 
Moreover for each $m$ there exists a positive $T$ and a unitary $U$ such that $m=TU$, so one can choose a positive 
representative in each class. The remaining symmetry is the equivalence 
\beas 
T \sim O T O^T , \qquad\qquad O \in O(2) 
\eeas
Now one can use a gauge fixing of the form 
\beas 
\mbox{Re} \, T_{12} = 0 
\eeas 
and employ the Faddeev--Popov method: The invariant measure on the space of selfadjoint matrices is known to be 
\beas 
{\rm d}   H ={\rm d}  p \, {\rm d} q \, {\rm d} \mbox{Re} (z) \,  {\rm d} \mbox{Im} (z) \qquad \mbox{for} \qquad 
H = \left(  \begin{array}{cc} p & z \\ \bar{z} & q \end{array} \right) \qquad\qquad p,q \in \R , z \in \C 
\eeas 
Starting from that fact, we calculate the invariant measure on the space of matrices $C$ of the form as in (\ref{DundC}), 
which is given by the well-known formula 
\beas 
{\cal Z} = {\cal N} \int\limits_{H \ge 0} {\rm d} H \, \delta ( \mbox{Re} H_{12} ) \, \Delta_{FP} e^{-S(H)} 
\eeas 
where 
\beas 
\Delta_{FP}^{-1} = \int {\rm d} \alpha \, \delta ( \mbox{Re} H_{12}^\alpha ) 
\eeas 
and $H_{12}^\alpha$ is the 12-element after the rotation of $H$ about $\alpha$ which is given by formula (\ref{ReT12}). The calculation of 
\beas 
\Delta_{FP}^{-1} = \int {\rm d} \alpha \, \delta \left( (q-p) \sin \alpha \cos \alpha + (\cos^2 \alpha - \sin^2 \alpha ) \mbox{Re} z  \right) 
\eeas 
can then be carried out using the formula 
\beas 
\delta (f(x)) = \sum_i \frac{1}{|f'(x_i)|} \delta (x-x_i) \quad \mbox{summing over the zeros} \, x_i \, \mbox{of} \, f 
\eeas
Finally one gets the (surprisingly simple) result 
\beas 
\Delta_{FP} (H) = \sqrt{(p-q)^2 + 4 ( \mbox{Re} z)^2 } 
\eeas 
Now writing everything in terms of matrices $C$, i.e. taking into account positivity and the condition $a \ge b \ge 0$, one finally gets for the invariant integral 
\beas 
\int f(C) {\rm d} C = \int\limits_0^\infty  {\rm d} a \int\limits_0^a {\rm d} b \int\limits_{-\sqrt{ab}}^{\sqrt{ab}} {\rm d} c f(C) 
\eeas 
and for the corresponding state sum -- with boundary condition (b.c.) $H \ge 0, \, H_{11} \ge H_{22} \ge 0$   
\beas 
{\cal Z} & = &  {\cal N}   \int\limits_{(b.c.)} {\rm d} H \delta ( \mbox{Re} H_{12} ) \Delta_{FP} e^{-S(H)} \\  
         & = & {\cal N} \int\limits_0^\infty {\rm d} a \int\limits_0^a {\rm d} b \int\limits_{-\sqrt{ab}}^{\sqrt{ab}} {\rm d} c \sqrt{ (a-b)^2} e^{-S(a,b,c)}   
\eeas 
One can see from the last expression that (even for the free classical action $S(D) = t_1 {\rm tr} D^2$) the strongest 
contribution does not stem from the configuration $C=0 \quad (\iff D = 0)$ and accordingly the vacuum expectation value for the distance 
remains finite: \\
The distance in this example is given by the inverse of the biggest
eigenvalue of $C$, i.e. 
\beas 
d(1,2) & = & \left( \frac{a+b}{2} + \sqrt{ \left( \frac{a-b}{2} \right)^2
+ c^2} \right)^{-1} 
\eeas 
For the action  
\beas 
S = \frac{1}{2} \tr D^2 & = & 2 (a^2 +b^2 + 2c^2) 
\eeas 
we therefore get 
\beas 
\langle d(1,2) \rangle & = & {\cal N} \int\limits_0^\infty {\rm d} a
\int\limits_0^a {\rm d} b \int\limits_{-\sqrt{ab}}^{\sqrt{ab}} {\rm d} c
\, \frac{(a-b)}{a+b+\sqrt{(a-b)^2+4c^2}} e^{-2(a^2+b^2+2c^2)}   
\eeas 
All constants are absorbed in ${\cal N}$ from now on. To show finiteness
we use the fact that 
\beas 
\frac{1}{a+b+\sqrt{(a-b)^2 +4c^2}} e^{-2(a^2+b^2+2c^2)}   & \le &
\frac{1}{2a} e^{-2(a^2+b^2)} 
\eeas 
(remember $a \ge b$ in the integration domain). This leads to 
\beas 
\langle d(1,2) \rangle & \le & {\cal N} \int\limits_0^\infty {\rm d} a
\int\limits_0^a {\rm d} b \int\limits_0^{\sqrt{ab}} {\rm d} c \,
\frac{a-b}{a} \, e^{-2(a^2+b^2)} 
\eeas 
The integration over $c$ gives $\sqrt{ab}$, so we have 
\beas 
\langle d(1,2) \rangle & \le & {\cal N} \int\limits_0^\infty {\rm d} a
\int\limits_0^a {\rm d} b \, \sqrt{ab} \, (1- \frac{b}{a} ) \,
e^{-2(a^2+b^2)} \\ 
                       & = & {\cal N} \int\limits_0^\infty {\rm d} a
\int\limits_0^a {\rm d} b \, \sqrt{ab} \,  e^{-2(a^2+b^2)} - {\cal N}
\int\limits_0^\infty {\rm d} a \int\limits_0^a {\rm d} b \,
\sqrt{\frac{b^3}{a}} \, e^{-2(a^2+b^2)} 
\eeas 
For the modulus we therefore get the estimate 
\beas 
| \langle d(1,2) \rangle | & \le & {\cal N} \int\limits_0^\infty {\rm d} a
\, \sqrt{a} \, e^{-2a^2} \int\limits_0^\infty {\rm d} b \, \sqrt{b} \,
e^{-2b^2} + {\cal N} \int\limits_0^\infty {\rm d} a \, \frac{1}{\sqrt{a}}
e^{-2a^2} \int\limits_0^\infty {\rm d} b \, \sqrt{b^3} \, e^{-2b^2} \\ 
    & < & \infty 
\eeas  
which is due to the fact that 
\beas 
\int\limits_0^\infty x^n \, e^{-k x^2} \, {\rm d} x <  \infty \quad
\mbox{for} \quad n > -1 
\eeas 
This shows finiteness of $\langle d(1,2) \rangle$.\\ \\  

The next example illustrates an effect of broken spectral invariance: 

\begin{example} 
{\rm For}  
\beas 
q = \left( \begin{array}{cc}  1 & -1 \\ -1 & 1 \end{array} \right) 
\eeas 
\end{example} 
\noindent 
(see example \ref{beisp4.2}) we had found the following expression for the most general Dirac operator: 
\beas 
D = \left( \begin{array}{cccc} 0 & m & \bar{m} & 0 \\ \bar{m} & 0 & 0 & \mu \\ m & 0 & 0 & \bar{\mu} \\ 0 & \bar{\mu} & \mu & 0 \end{array} \right) \quad m , \mu \in \C
\eeas 
and one of the two complex numbers (say $m$) could be chosen real positive. For the distance we had  
\beas 
d(1,2) = \frac{1}{\max \{ m , |\mu | \} } 
\eeas 
whereas the eigenvalues of the Dirac operator are given by 
\beas \label{eigvdo}
\lambda_{\pm}^2 = m^2 + |\mu|^2 \pm |m^2 + \mu^2 |
\eeas 
and depend on the phase of $\mu$. It is therefore impossible to express the distance in terms of the eigenvalues (this was already discussed in example \ref{dreieins}). We shall therefore only seek for a measure that is invariant under diffeomorphisms -- which is trivial here --
but not under all unitaries on $\CH$.\\
Such a diffeomorphism invariant state sum for $S(D) = \frac{t}{4}$ tr $D^2 = t(m^2 + |\mu|^2)$ is then given by 
\beas 
{\cal Z} & = & {\cal N} \int\limits_0^\infty {\rm d} m \int {\rm d} \mu  \, {\rm d} \bar{\mu}  \, e^{-t(m^2+|\mu|^2)} \label{Zdieerste} \\ 
         & = & 2 \pi {\cal N} \int\limits_0^\infty {\rm d} m \, {\rm d} r \, r e^{-t(m^2 + r^2 )} \\ 
         & = & 2 \pi {\cal N} \int\limits_0^\infty {\rm d} m \, {\rm d} r e^{-W(m,r)} 
\eeas
where $W(m,r):= t(m^2 + r^2) - \ln r$. If we use the estimate 
\beas 
\sqrt{m^4 + |\mu|^4 + 2m^2 |\mu|^2 \cos (2\varphi)} \ge \sqrt{(m^2 - |\mu|^2 )^2} 
\eeas
we get 
\beas 
\lambda_-^2 & = & m^2 + |\mu|^2 - \sqrt{m^4 + |\mu|^4 + 2m^2 |\mu|^2 \cos (2\varphi)} \\ 
           & \le &  m^2 + |\mu|^2 - \sqrt{(m^2 - |\mu|^2 )^2} \\ 
           & = & 2 \min \{ m^2 , |\mu|^2 \}  
\eeas
and thus 
\beas 
|\lambda_+ | - |\lambda_-| \ge \sqrt{2} | m - |\mu|| 
\eeas 
Using this result  we can then calculate 
\beas 
\langle |\lambda_+ | - |\lambda_-| \rangle & \ge & \sqrt{2} \langle |m- |\mu| | \rangle \\ 
  & = & {\cal N} \int\limits_0^\infty {\rm d} m {\rm d}r \, r |m-r|e^{-(m^2 + r^2)} \label{nichtspecinv} \\ 
  & = & \frac{ {\cal N} }{4t^2} > 0 
\eeas 
showing that the vacuum expectation value of the two quantized eigenvalues is different. 
But this should not be the case for a spectral invariant quantum theory, because of the following observation (already discussed in the introduction): If 
\beas 
S(D^2) = \sum_i P(\lambda_i^2 ) 
\eeas
is spectral invariant and if the polynomial $P(\lambda)$ has a unique extremum, then all the eigenvalues of $D^2$ at the 
extremal point of $S$ are identical (which follows directly from spectral invariance, because the groundstate must be 
invariant especially under permutations of the eigenvalues). \\ 
It follows from the small calculation above that the unitary transformations that correspond to these transformations 
are not represented in the quantized theory in such a way that the groundstate is invariant. 
Clearly, this violation of spectral invariance can be traced back to the fact that the measure we used is not spectral invariant, 
which can be clarified by taking a look at the effective action 
\beas 
W(m,r):= t(m^2 + r^2) - \ln r.
\eeas 
Its minimum lies at the point $r=\frac{1}{\sqrt{2t}}$ and not at $r=0$.  \\ \\ 
The next example deals with the triple which was already defined in example \ref{2-1-10}. 
\begin{example} 
\beas 
q = \left( \begin{array}{cc}  2 & -1 \\ -1 & 0 \end{array} \right)  
\eeas
\end{example} \noindent
Dirac operators are given by 
\beas 
D = \left( \begin{array}{cccc} 0 & 0 & m_1 & \bar{m_1} \\ 
                               0 & 0 & m_2 & \bar{m_2} \\ 
                               \bar{m_1} & \bar{m_2} & 0 & 0  \\ 
                               m_1 & m_2 & 0 & 0  \end{array} \right) \qquad \qquad m_1 , m_2 \in \C 
\eeas
and representatives of the equivalence classes are characterized by 
\beas 
m_1 = \frac{\rho}{\sqrt{2}} \quad m_2 = \frac{\rho}{\sqrt{2}} e^{i \psi} 
\eeas  
The eigenvalues of $D$ in this parametrization are given by 
\bea
\lambda_+^2 & = & \frac{\rho^2}{2} \left( 2 + \sqrt{2+2 \cos (2 \psi)} \right) \label{lambdaplusquadrat} \\ 
\lambda_-^2 & = & \frac{\rho^2}{2} \left( 2 - \sqrt{2+2 \cos (2 \psi)} \right)  \label{lambdaminusquadrat} 
\eea
\noindent
So for 
\beas 
S(D) = \frac{t}{4} \tr D^2 = t \rho^2 
\eeas 
one can for example calculate 
\bea 
\langle \lambda_+^2 - \lambda_-^2 \rangle = \left\langle \rho^2 \sqrt{2+2
\cos (2 \psi)} \right\rangle 
\eea 
with the (obvious) measure 
\bea \label{dasmassrhopsi}
\int {\rm d} \rho \int {\rm d} \psi 
\eea 
The only thing we need to take care of is the interval of integration for
the variable $\psi$ which is $[0, \frac{\pi}{2} ]$ due to theorem
\ref{RmUtheorem}. So for this case we get 
\bea \label{confusingformel} 
\int\limits_0^\infty {\rm d} \rho \int\limits_0^{\frac{\pi}{2}}  {\rm d}
\psi \rho^2 \sqrt{2+2 \cos (2 \psi)} e^{-t \rho^2} = \sqrt{
\frac{\pi}{4t^3}} > 0 
\eea
Thus, also in this example we observe the effect of spontaneously broken invariance.
Unlike the previous example, the measure we used here is, however, spectral invariant: 
\\
Remember that (if there are four independent eigenvalues of $D$) a spectral invariant measure must be of the form 
\beas 
f (\lambda_1 ,\lambda_2 , \lambda_3, \lambda_4 ) \, {\rm d} \lambda_1 \,
{\rm d} \lambda_2 \, {\rm d} \lambda_3 \, {\rm d}\lambda_4
\eeas 
where $f$ is totally symmetric with respect to permutations of its
arguments. In our case the eigenvalues of the Dirac operator are not
independent: $\lambda_1 = - \lambda_2 , \lambda_3 = - \lambda_4$ so one
can as well choose the invariant measure as 
\beas 
f (\lambda_1 , \lambda_3 ) \, {\rm d} \lambda_1 \, {\rm d} \lambda_3 
\eeas 
(in the integral $\lambda_1$ and $\lambda_3$ are now running from $0$ to
$\infty$). To show invariance of (\ref{dasmassrhopsi}) we first prove the
following 
\begin{lema} \label{spektralinvlemma} 
If the function $f$ has the properties $|f( \lambda_+ , \lambda_- )| = | f
(\lambda_- , \lambda_+ )| \quad \forall \lambda_+ , \lambda_-$ and
$f(\lambda_+ , \lambda_- ) \ge 0$ if $ \lambda_+ \ge \lambda_-$ then the
expression $\int\limits_0^\infty {\rm d} \lambda_+
\int\limits_0^{\lambda_+} {\rm d} \lambda_- \, f (\lambda_+ , \lambda_-
) $ is spectral invariant. 
\end{lema} 
PROOF: According to the assumption (the symmetry of $|f|$) the expression 
\beas 
\int\limits_0^\infty {\rm d} \lambda_+ \int\limits_0^\infty  {\rm d}
\lambda_- \, | f(\lambda_+ , \lambda_- ) |  
\eeas 
is spectral invariant. If one splits the domain of integration as follows 
\beas 
\int\limits_0^\infty {\rm d} \lambda_+ \int\limits_0^\infty {\rm d}
\lambda_-    =  \int\limits_0^\infty {\rm d} \lambda_+
\int\limits_0^{\lambda_+} {\rm d} \lambda_-  + \int\limits_0^\infty {\rm
d} \lambda_- \int\limits_0^{\lambda_-} {\rm d} \lambda_+ 
\eeas 
one gets 
\beas 
\int\limits_0^\infty {\rm d} \lambda_+ \int\limits_0^\infty {\rm d}
\lambda_-  \, |f(\lambda_+ , \lambda_- )| & = & \int\limits_0^\infty {\rm
d} \lambda_+ \int\limits_0^{\lambda_+} {\rm d} \lambda_- |f(\lambda_+,
\lambda_-)|   + \int\limits_0^\infty {\rm d} \lambda_-
\int\limits_0^{\lambda_-} {\rm d} \lambda_+ |f(\lambda_+, \lambda_-)| 
\eeas 
After renaming $\lambda_+ \leftrightarrow \lambda_-$ and using the
symmetry of $|f|$ one gets equality of the two terms on the r.h.s. and
finally 
\beas 
\int\limits_0^\infty {\rm d} \lambda_+ \int\limits_0^\infty {\rm d}
\lambda_-  \, |f(\lambda_+ , \lambda_- ) & = & 2 \int\limits_0^\infty {\rm
d} \lambda_+ \int\limits_0^{\lambda_+} {\rm d} \lambda_- |f(\lambda_+,
\lambda_-)   
\eeas 
Now we can use the property $|f(\lambda_+ , \lambda_-) | = f(\lambda_+ ,
\lambda_-) \, \forall \lambda_+ \ge \lambda_-$ to see that 
\beas    
\int\limits_0^\infty {\rm d} \lambda_+ \int\limits_0^{\lambda_+} {\rm d}
\lambda_- \, f(\lambda_+ , \lambda_-) & = & \frac{1}{2}
\int\limits_0^\infty {\rm d} \lambda_+ \int\limits_0^\infty  {\rm d}
\lambda_- \, | f(\lambda_+ , \lambda_-) | 
\eeas 
is indeed spectral invariant. \hfill \checkmark \\ \\ 

Let's now check, whether the measure (\ref{dasmassrhopsi}) fulfills the
requirements of lemma \ref{spektralinvlemma}. 
Summing (\ref{lambdaplusquadrat}) and (\ref{lambdaminusquadrat}) leads to 
\beas 
\lambda_+^2 + \lambda_-^2 & = & 2 \rho^2 \\ 
\Rightarrow \rho & = & \sqrt{ \frac{1}{2} (\lambda_+^2 + \lambda_-^2 )} 
\eeas 
Taking the difference of the two expressions gives 
\beas 
\lambda_+^2 - \lambda_-^2 & = & \rho^2 \sqrt{2+2 \cos (2 \psi)} \\ 
                          & = & \frac{1}{2} (\lambda_+^2 + \lambda_-^2
) \sqrt{2+2 \cos (2 \psi)} \\ 
& & \\
\Rightarrow \cos (2 \psi )  & = & 2 \left( \frac{\lambda_+^2 -
\lambda_-^2}{\lambda_+^2 + \lambda_-^2 } \right)^2  -1 \\ 
& & \\
\Rightarrow \psi & = & \frac{1}{2} \arccos \left( 2
\left(\frac{\lambda_+^2 - \lambda_-^2}{\lambda_+^2 + \lambda_-^2 }
\right)^2  -1 \right)  
\eeas 
In the following we put $x:= \lambda_+ , y:= \lambda_-$ to alleviate
notation. So we have 
\beas 
\rho & = & \sqrt{ \frac{1}{2} (x^2 + y^2)} \\ 
\psi & = & \frac{1}{2} \arccos \left( 2 \left(\frac{x^2 - y^2}{x^2 + y^2 }
\right)^2  -1 \right)  
\eeas 
The measure transforms according to 
\beas 
{\rm d} \rho & = & \frac{\partial \rho}{\partial x} {\rm d} x +
\frac{\partial \rho}{\partial y} {\rm d} y \\ 
{\rm d} \psi & = & \frac{\partial \psi}{\partial x} {\rm d} x +
\frac{\partial \psi}{\partial y} {\rm d} y \\ 
\Rightarrow {\rm d} \rho \, {\rm d} \psi & = & \underbrace{ \left(
\frac{\partial \rho}{\partial x} \frac{\partial \psi}{\partial y} -
\frac{\partial \rho}{\partial y} \frac{\partial \psi}{\partial x}
\right) }_{=: J(x,y)} {\rm d} x {\rm d}  y 
\eeas 
Calculating $J$ leads to 
\beas 
J(x,y) = \frac{4xy(x^2-y^2)}{(x^2+y^2)^2 \sqrt{ 2 \frac{(x^2 - y^2
)^2}{x^2+y^2} \left[ 1 - \left( \frac{x^2-y^2}{x^2+y^2} \right)^2 \right]
} } = \frac{\sqrt{2}}{\sqrt{x^2+y^2}} \sign (x^2-y^2) 
\eeas 
As one can see, $J$ has the properties that are postulated in lemma
\ref{spektralinvlemma} and so the integral 
\beas 
\int\limits_0^\infty {\rm d} \rho \int\limits_0^{\frac{\pi}{2}} {\rm d}
\psi =  \int\limits_0^\infty {\rm d} \lambda_+ \int\limits_0^{\lambda_+}
{\rm d} \lambda_- \, J(\lambda_+ , \lambda_- ) 
\eeas  
is spectral invariant. 
Thus there is a less obvious loss of spectral invariance in the transition from
classical to quantized theory in this example.

\newpage

\section{Discussion/outlook} 
In this paper we entered the subject of quantizing finite dimensional spectral triples only superficially. In
fact, we mentioned only few points concerning only two points. However, from such a baby toy model one should not
expect more than an incomplete illustration.\\
In particular, the reader might have missed the usual folkloristic results about the Planck length, i.e. a minimal measurable
distance of (the) two points. We have, actually,  not been able to obtain such results in our models.
For the simplest example \ref{daseinfachste}, when $D$ has only one eigenvalue, and for the gaussian measure, the vacuum expectation 
value of the distance is given by
\[  \langle 0 | \,d(1,2) \, | 0\rangle = {\cal N} \int\limits_0^\infty \, {\rm d} \lambda \,\, \frac{1}{\lambda}\, e^{-t\lambda^2}\,. \]  
For an arbitrary state 
\[ | f \rangle = f(\lambda) |0\rangle \qquad\qquad\qquad 
 \langle f| f \rangle = {\cal N} \int\limits_0^\infty \, {\rm d} \lambda \,\, |f|^2 \, e^{-t\lambda^2} = 1  \]
with some suitable function $f$, one then easily verifies that the expectation value
\[  \langle f|\, d(1,2)\,| f \rangle = {\cal N} \int\limits_0^\infty \, {\rm d} \lambda \,\, \frac{|f|^2}{\lambda} \, e^{-t\lambda^2}\]
is only bounded from below by zero: Consider, for instance,  
$| f(\lambda)|^2 = \frac{e^{t\lambda_0^2}}{\cal N}\, \delta (\lambda-\lambda_0)$, in which case 
$\langle f|\, d(1,2)\,| f \rangle = \frac{1}{\lambda_0}$. 
\\
However, such arguments do not appear too compelling (to us) in view of the severe problem of lacking time. 
As mentioned in the introduction, models based on algebras 
\[ \CA = C^\infty_0(\R) \otimes \CA_F  \]
where $\CA_F$ denotes any finite dimensional $C^*$-algebra, while $C^\infty_0(\R)$ is interpreted as functions on the time axis,
are currently under construction. It is our ultimate aim in this project, to investigate models of this type. It is then
possible to meaningfully define classical, canonical transformations -- under which a diffeomorphism invariant measure 
for the quantization could be invariant -- and so on. Even more so, one can then dream of approximating spacetime
by such a model, or more precisely: space by $\CA_F$. But all that is music of the future (but not of the past).
\\
That does not mean, however that it is in vain to study models based on finite dimensional spectral triples.
First of all, solving the technical problems they pose, is an unavoidable preparation before dealing with the more complicated 
models where time is included. Secondly, though the ``physical'' interpretation of these models is not clear from the start,
it is also not clear that it doesn't exist. Moreover these systems show some remarkable effects, for which we would like to gain some 
intuition. For example, we have seen that the vacuum expectation value of the distance of the two points is infinite
when $D$ has only one independent eigenvalue, but comes out being finite if there are more eigenvalues of $D$. In the example we presented, 
this seems to be related to the spontaneous breaking of spectral invariance, but in other examples this is not the case:
In a forthcoming paper we shall for instance present an example, where some ``distance-like'' observable is given as
$d = \frac{1}{\max \{\lambda_1,\lambda_2\} }$. Here $\lambda_1,\lambda_2$ denote the two independent eigenvalues of $D$ --
if there were only one eigenvalue, $d$ would equal the inverse of this value -- and the expectation value is given as
\[ \langle d \rangle = {\cal N} \int\limits_0^\infty \int\limits_0^\infty \, {\rm d}\lambda_1{\rm d}\lambda_2 \,\,
\frac{1}{\max \{\lambda_1,\lambda_2\} } e^{-t(\lambda_1^2 + \lambda_2^2)} .
\]
It is a nice exercise to compute this {\em finite} expectation value  \cite{diss}. 
\\
Last not least, as they lead to rather unusual matrix models, the quantization of finite dimensional spectral triples
might be interesting enough by itself. In this respect, it is particularly challenging to study the various possible
continuum limits these systems offer:
\\ 
Most obviously one could consider $N$-point spaces and eventually the continuum-limit $N \to \infty$ which corresponds to a lattice. 
Secondly, one might study proper noncommutative examples (which is our aim in the subsequent paper), 
moreover one can consider limits $n_i \to \infty$ in the decomposition of the algebra 
\beas 
A = \bigoplus_{i=1}^N M_{n_i} (\C) .
\eeas  
But one could also study limits where some or all entries of the intersection form $q$ are sent to infinity. All that, of course,
requires a much more systematic construction of path integrals for (as generic as possible) finite dimensional spectral
triples, and this will therefore be the subject of our subsequent paper.

\newpage
\begin{appendix}
\section{A proof of the canonical form of $J$ }
As the proofs in the literature of the canonical form of $J$ that we have stated in section \ref{review}
are not completely compelling, we would like to present here another proof of this (nevertheless correct) statement.
\\
Before giving our new proof we shall briefly describe at which point it essentially improves on
the literature: \\
As an antiunitary operator $J$ is of the form $J = K U$, where $K$ denotes the antilinear operator of complex conjugation,
while $U$ is unitary. 
All the proofs in the literature use the fact that $U$ can be diagonalized. Exploiting the antilinearity of $J$, it is then rather easy
to get rid of the eigenvalues of $U$ by a suitable redefinition of the basis.  \\
It is however not shown that one can also diagonalize the combination $K U$, i.e. that there exists a unitary matrix $R$ such that
\[ R\,\, KU \,\, R^* = K\,\, \bar{R} U R^* = K U_d \]
where $U_d$ is diagonal. We shall prove that one can in fact find a real (orthogonal) matrix $R$ doing the job.
\begin{lemma} \label{JBasis} 
There always exists an othornormal basis $\{ v_1 , \ldots , v_n  \}$ of
$\CH_{kl}$ and $\{ w_1 , \ldots , w_n \}$ of $\CH_{lk}$ respectively, such that 
for $k \ne l$  one has \[J v_i = w_i\] and for $k=l$  (with $v_i
= x \otimes y \otimes z$) it is
\begin{equation} 
J v_i = J (x \otimes y \otimes z) = \bar{z} \otimes \bar{y} \otimes
\bar{x} .
\end{equation} 
Thus $J$ essentially interchanges the basis vectors of the different subspaces of $\CH$.
\end{lemma} 
PROOF: First of all note that $J$ defined by  
\begin{equation} \label{defvonJ} 
J ( u \otimes v \otimes w ) = \bar{w} \otimes \bar{v} \otimes \bar{u} 
\end{equation} 
fulfills the axioms $J^2 =1$ and $[\pi (a) , J \pi (b) J ]=0$:
($J^2 = 1$ is obvious) 
\begin{eqnarray*} 
\pi (a) J \pi (b) J ( u \otimes v \otimes w  ) & = & \pi (a) J \pi (b) (
\bar{w} \otimes \bar{v} \otimes \bar{u} ) \\ 
                                               & = & \pi (a) J ( (\pi
(b)  \bar{w} ) \otimes \bar{v} \otimes \bar{u} ) \\ 
                                               & = & \pi (a) ( u \otimes v
\otimes \overline{\pi (b) \bar{w}  } )  \\ 
                                               & = & ( \pi (a) u ) \otimes
v \otimes \overline{ \pi (b) \bar{w} } 
\end{eqnarray*} 
\begin{eqnarray*} 
J \pi (b) J \pi (a) (  u \otimes v \otimes w ) & = & J \pi (b) J ( (\pi
(a) u) \otimes v \otimes w) \\ 
                                               & = & J \pi (b) ( \bar{w}
\otimes \bar{v} \otimes \overline{\pi (a) u} ) \\ 
                                               & = & J ( (\pi (b) \bar{w}
) \otimes \bar{v} \otimes \overline{\pi (a) u} ) \\ 
                                               & = & (\pi (a) u) \otimes v
\otimes  \overline{ \pi (b) \bar{w} }.  
\end{eqnarray*} 
Now, let $\tilde{J}$ be another antiunitary map with the properties $\tilde{J}^2 =
1 $ and $\pi^o (a)  = \tilde{J} \pi (a^*) \tilde{J}$. Then  $J\circ
\tilde{J}$ is an invertible map that commutes with both representations of $\CA$, 
\begin{equation*} 
\, [J \circ \tilde{J} , \pi (a) ] = [J \circ \tilde{J} , \pi^o (a) ]
= 0 
\end{equation*} 
since: 
\begin{eqnarray*} 
J \circ \tilde{J} \pi (a) & = & J \circ \tilde{J} \pi (a) \tilde{J}^2  = J
\circ \pi^o (a^*) \circ \tilde{J} \\ 
                          & = & J \circ J \circ \pi (a) \circ J \circ
\tilde{J}  =  \pi (a) J \circ \tilde{J} 
\end{eqnarray*} 
and analogously for $\pi^o$. 
Thus $J \circ \tilde{J}$ is of the form 
\begin{equation} \label{formvonjjschlange} 
J \circ \tilde{J} = {\bf 1} \otimes j \otimes  {\bf 1}.
\end{equation} 
As $J \circ \tilde{J}$ is unitary: 
\begin{equation*} 
\langle \psi , \phi \rangle = \langle \tilde{J} \phi , \tilde{J} \psi
\rangle = \langle J \circ \tilde{J} \psi , J \circ \tilde{J} \phi \rangle 
\end{equation*} 
also $j$ is unitary.  Consider the following two cases: \\ 
{\em Case 1}: $k \ne l$.\\ 
Choose an arbitrary basis $\{ v_1 , \ldots , v_n \}$
in $\mathbb{C}^{r_{kl}} $ and define the basis in
$\mathbb{C}^{r_{lk}}$ through
\begin{equation*} 
w_i := J \circ (1 \otimes j \otimes 1) v_i.  
\end{equation*} 
Since $\tilde{J} = J \circ (1 \otimes j \otimes 1) $ it then immediately follows that
$w_i = \tilde{J} v_i$ holds. \\ 
{\em Case 2}: $k=l$. \\
Note that $J^2 = \tilde{J}^2 = 1$, which implies $j \bar{j}
=1$:
\begin{eqnarray*} 
(\ref{formvonjjschlange}) \Rightarrow \tilde{J} & = & J \circ (1 \otimes j
\otimes 1) \\ 
 \Rightarrow \tilde{J}^2 & = & J \circ (1 \otimes j \otimes 1) \circ J
\circ (1 \otimes j \otimes 1) \\ 
 & = & (1 \otimes \bar{j} \otimes 1) \circ J^2 \circ (1 \otimes j \otimes
1) \quad \mbox{according to (\ref{defvonJ})} \\ 
 & = & (1 \otimes  \bar{j} j \otimes 1). 
\end{eqnarray*} 
Thus $j$ fulfills the relations $j^* = j^{-1}$ and $\bar{j} =j^{-1}$. 
To such a matrix $j$ there always exists a unitary matrix
$u$ with the property 
\bea \label{zerlegungvonj} 
j = u^T u 
\eea
{\em Proof of formula (\ref{zerlegungvonj})}:\\
Consider $j \in M_n (\C)$. We prove the statement by induction on $n$.
\\
The case $n=1$ is trivial: For $j=e^{i
\varphi}$ one can simply take $u = e^{i \varphi /2}$. \\ 
Suppose now the statement has been proven for all $k=1,\ldots , n$ and let $j \in
M_{n+1} (\C)$. 
Since $j$ is unitary, its eigenvalues $\lambda_1 ,
\ldots , \lambda_{n+1}$ are complex with modulus $1$. Since $\bar{j} = j^{-1}$
the following is true: \\
If $\psi$ is an eigenvector of $j$ for the eigenvalue $\lambda$, 
then also its complex conjugate vector $\bar{\psi}$ is an eigenvector for the same  eigenvalue $\lambda$. 
To see this, note that a matrix and is inverse do have the same eigenvectors, and thus also
$j$ and $\bar{j}$ do have the same eigenvectors (which is not true in general).
Hence one has the equation (recall $\lambda^{-1}=\bar{\lambda}$ )
\[ \bar{j} \psi = \bar{\lambda} \psi \]
which after complex conjugation shows that $\bar{\psi}$ is an eigenvector of $j$ for the eigenvalue
$\lambda$.\\ 
Now if $j$ had only one eigenvalue $\lambda$, being ($n+1$)-times degenerate, 
it would be proportional to the identy $j = \lambda {\bf 1}$. So in such a case the existence of $u$ with $j=u^T u$ would be obvious. \\ 
Accordingly we can suppose in the following that $j$ has at least two different eigenvalues. Let $\lambda$ be
one of these eigenvalues, $E_\lambda$ be the corresponding eigenspace. By assumption its dimension
$k$ can at most equal $n$.  Let $\{ \psi_1 , \ldots ,
\psi_k \}$ denote an orthonormal basis of $E_\lambda$. Then also the complex conjugate vectors belong 
to $E_\lambda$ and one easily verifies that $\{ \bar{\psi}_1 ,
\ldots , \bar{\psi}_k \}$ is another orthonormal basis of $E_\lambda$. 
The interchange between these two (possibly different) bases is described by a unitary matrix $a$: 
\bea\label{basiswechsela}
\bar{\psi}_i = a_{ji} \psi_j 
\eea 
(with summation over the index $j$).
But then by complex conjugation of (\ref{basiswechsela}) it follows that $\psi_i =
\bar{a}_{ji} \bar{\psi}_j $ and hence  $\psi_j = \bar{a}_{kj}
\bar{\psi}_k$. Plugging this expression into
(\ref{basiswechsela}), we conclude :  
\beas 
\bar{\psi}_i & = & a_{ji} \bar{a}_{kj} \bar{\psi}_k \\ 
\Rightarrow \bar{a}_{kj} a_{ji} & = & \delta_{ki} \qquad\qquad \mbox{(since} \quad
\{ \bar{\psi}_i \} \quad \mbox{is ONB)} \\ 
\Rightarrow \bar{a} & = & a^{-1} .
\eeas 
Conclusively the  matrix $a$ fulfills the hypothesis of the induction and hence it can be written as  
\[ a=b b^T \iff b_{ij} b_{kj} = a_{ik}.\] If we now define 
a new ONB of $E_\lambda$ through 
\beas 
\phi_i := b_{ji} \psi_j 
\eeas 
then it follows: 
\beas 
\bar{\phi}_i & = & \bar{b}_{ji} \bar{\psi}_j \\ 
&=  &\bar{b}_{ji} a_{kj} \psi_k \\ 
             & = & \bar{b}_{ji} b_{kl} b_{jl} \psi_k \\
& =  & b_{kl} \underbrace{b^*_{ij} b_{jl}}_{\delta_{il}}  \psi_k \\ 
             & = & b_{ki} \psi_k \\
& =  & \phi_i  
\eeas 
Thus $\{ \phi_1 , \ldots , \phi_k \}$ is a {\em real} orthonormal basis of 
$E_\lambda$. Since this is true for any eigenspace of $j$, we conclude that on can find a complete
set of real, orthonormal eigenvectors of $j$. Hence $j$ can be diagonalized by a (real) orthogonal matrix
$R$: 
\beas 
R j R^T = j_d = \mbox{diag} (e^{i\rho_1} , \ldots , e^{i\rho_{n+1}} )
\eeas 
Set $u = \mbox{diag} (e^{i\rho_1/2} , \ldots , e^{i\rho_{n+1}/2}
) R$. Then one checks $j = u^T u$ and this proves (\ref{zerlegungvonj}) . \\ 
\\
As a consequence of the decomposition $j= u^T u$ it then follows that 
\begin{eqnarray*} 
(1 \otimes u \otimes 1) \circ \tilde{J} \circ (1 \otimes  u^* \otimes 1) &
= & (1 \otimes u \otimes 1) \circ  J \circ (1 \otimes j \otimes 1)  \circ
(1 \otimes u^* \otimes 1) \\ 
 & = & J \circ (1\otimes \bar{u} j u^* \otimes 1) .
\end{eqnarray*} 
But $\bar{u} j u^* = \bar{u} u^T u u^* =1$, and therefore  
\beas 
(1 \otimes u \otimes 1) \circ \tilde{J} \circ (1 \otimes  u^* \otimes 1) =
J .
\eeas 
So changing the basis with the help of $u$ will cast $\tilde{J}$ in the desired form.
\hfill \checkmark
\end{appendix}

\end{document}